\documentclass[12pt]{article}
\usepackage{mathptmx}
\usepackage{amsmath}
\usepackage{latexsym}
\usepackage{amsfonts}
\usepackage[normalem]{ulem}
\usepackage{soul}
\usepackage{array}
\usepackage{amssymb}
\usepackage{extarrows}
\usepackage{adjustbox}
\usepackage{physics}
\usepackage{textgreek}
\usepackage{steinmetz}
\usepackage{comment}
\usepackage[british]{babel}
\usepackage[style=british]{csquotes}
\usepackage[
    backend=biber,
    giveninits=false,
    style=ieee,
    sorting=none,
    doi=false
]{biblatex}
\usepackage{biblatex}
\DeclareFieldFormat{labelnumberwidth}{#1\adddot}
\DeclareFieldFormat{journaltitle}{#1}
\DeclareFieldFormat[article]{volume}{Vol\addnbspace #1}
\DeclareFieldFormat[article]{number}{No \addnbspace #1}
\DeclareFieldFormat[article]{pages}{pp \addnbspace #1}

\usepackage{subfig}
\usepackage{wrapfig}
\usepackage{wasysym}
\usepackage{enumitem}
\usepackage{adjustbox}
\usepackage{ragged2e}
\usepackage[svgnames,table]{xcolor}
\usepackage{tikz}
\usepackage{longtable}
\usepackage{changepage}
\usepackage{setspace}
\usepackage{hhline}
\usepackage{multicol}
\usepackage{tabto}
\usepackage{float}
\usepackage{multirow}
\usepackage{makecell}
\usepackage{fancyhdr}
\usepackage[toc,page]{appendix}
\usepackage[hidelinks]{hyperref}
\usetikzlibrary{shapes.symbols,shapes.geometric,shadows,arrows.meta}
\tikzset{>={Latex[width=1.5mm,length=2mm]}}
\usepackage{flowchart}\usepackage[paperheight=11.69in,paperwidth=8.27in,left=0.98in,right=0.98in,top=1.48in,bottom=1.48in,headheight=1in]{geometry}
\usepackage[T1]{fontenc}

\usepackage{lineno}

\renewcommand{\_}{\kern-1.5pt\textunderscore\kern-1.5pt}

\usepackage{indentfirst}

\usepackage{secdot}

\usepackage{titlesec}
\titleformat*{\subsection}{\large\bfseries\itshape}
\titleformat*{\subsubsection}{\normalfont\itshape}

\usepackage{amsthm}
\newtheorem{theorem}{Theorem}
\theoremstyle{remark}
\newtheorem{remark}{Remark}

\begin{document}
\section*{\centering Old but Not Obsolete: Dimensional Analysis in Nondestructive Testing and Evaluation}

\vspace{\baselineskip}
\begin{Center}
Antonello Tamburrino\textsuperscript{1,2}, Alessandro Sardellitti\textsuperscript{1}, Filippo Milano\textsuperscript{1}, Vincenzo Mottola\textsuperscript{1}, Marco Laracca\textsuperscript{3},
 and Luigi Ferrigno\textsuperscript{1,4}
\end{Center}

\vspace{\baselineskip}
\begin{Center}
\textsuperscript{1}{D}ept. of Electrical and Information Engineering, University of Cassino and Southern Lazio,\\ 03043 Cassino (FR), Italy \\ e-mail: \{antonello.tamburrino, alessandro.sardellitti, filippo.milano, vincenzo.mottola, luigi.ferrigno\}@unicas.it
\end{Center}
\begin{Center}
\textsuperscript{2}{D}epartment of Electrical and Computer Engineering, Michigan State University, East Lansing, MI-48824, USA
\end{Center}
\begin{Center}
\textsuperscript{3}{D}ept. of Astronautics, Electrical and Energy Engineering, Sapienza University of Rome, \\ 00186 Rome, Italy \\
e-mail: marco.laracca@uniroma1.it
\end{Center}
\begin{Center}
\textsuperscript{4}{C}onsorzio Nazionale Interuniversitario per le Telecomunicazioni (CNIT), Italy
\end{Center}

\vspace{\baselineskip}
\begin{FlushLeft}
\textbf{Abstract} 
\end{FlushLeft}

This paper proposes dimensional analysis for solving Non--Destructive Testing \& Evaluation (NDT\&E) problems. This is the first time that this approach has been adopted in the framework of NDT\&E, and the paper ushers in the development of probes and methods for simultaneously estimating several parameters with a simple approach.

The most important theorem of dimensional analysis is Buckingham's \textpi \ theorem, based on the concept that the laws of the physics do not depend on the particular set of units chosen. At the core of this theorem is a systematic reduction in the number of variables describing a physical problem. This reduction is equal to $k$, the number of fundamental dimensions required to describe the variables of the  physical problem in its original setting. This makes the approach ideal when the number of variables in the physical problem is not much greater than $k$.

In this paper, we demonstrate the approach's effectiveness in the simple problem of simultaneously estimating the thickness and electrical conductivity of a conducting plate, via Eddy Current Testing. The approach is original, effective, efficient, and currently patent-pending. All the aspects, from theory to experimental validation, are provided, and it is proved that the proposed method achieves a very good degree of accuracy over a wide range of thicknesses and electrical conductivities. Moreover, the proposed method is compatible with the in-line and real-time estimation of thickness and electrical conductivity in an industrial environment.

\vspace{\baselineskip}
{\textbf{Keywords:} Dimensional analysis; Buckingham’s \textpi\ theorem; Non--Destructive Testing \& Evaluation (NDT\&E); Eddy Current Testing (ECT); Multi-parameter simultaneous estimation; Thickness estimation; Electrical conductivity estimation.}

\section{Introduction} \label{section:introduction} 
Problems related to Non--Destructive Testing and Evaluation (NDT\&E) generally involve several variables. Specifically, the outcome of an NDT\&E test depends on \textit{(i)} the parameters  describing the probe (geometry, materials, ...), \textit{(ii)} the physical and geometrical parameters of the sample being tested, \textit{(iii)} the geometrical parameters describing the position of the probe with respect to the sample being tested and \textit{(iv)} environmental factors. The number of variables involved and the correlated nature of these variables (e.g., excitation frequency and thickness estimation \cite{Sardellitti,Sardellitti1,Sardellitti2,Sardellitti3}) make an NDT\&E problem difficult to handle. 

To this end, a methodology that can systematically reduce the complexity of the analysis of NDT\&E problems by decreasing the number of variables involved, plays a very important role.

For instance, this reduction in the number of variables has a major impact when a physical problem is modelled via either a numerical or a machine learning approach {\cite{Machine_learning1,Machine_learning2,Machine_learning3,Bilicz,Miorelli1,Miorelli2,Miorelli3}. In both cases there is an exponential reduction in the number of numerical simulations required or the size of the training database.

To this purpose, dimensional analysis is a mathematical technique for analyzing problems involving physical quantities \cite{gibbi}.

Dimensional analysis can be used to simplify the analysis of complex equations by highlighting the fundamental quantities describing a problem. Specifically, by only analyzing the physical dimensions of the variables involved in an equation, it is possible to determine a smaller number of fundamental quantities describing the original problem. This simplifies the computation of the solution to the original problem \cite{GEORGI}. Dimensional analysis is commonly used in physics, engineering and other sciences to derive equations and verify experimental results  \cite{Curtis,Reddy,Ciulla}. The present paper is the first systematic study on the beauty and effectiveness of dimensional analysis, other than \cite{Sardellitti1} where dimensional analysis was merely applied in a thickness estimation problem. 

Within dimensional analysis, Buckingham’s \textpi\ theorem plays a key role. This theorem is grounded on the concept that the equations of physics cannot be affected by the choice of the physical quantities' units \cite{Buck1}. It states that any physical law can be written in terms of dimensionless groups and provides a procedure to identify these dimensionless parameters, also called \textpi\ groups. The key is that the number of \textpi \ groups is smaller than that of the original variables. For instance, if a physical problem is modelled by an equation of the type $g\left( q_1, \ldots, q_n \right)=0$ and the physical dimensions of the $q_i$s are expressed by a set of $k$ fundamental dimensions, then Buckingham's \textpi \ theorem allows the original physical problem to be cast as $G \left( \pi_1,\ldots,\pi_p \right)=0$, where $p=n-k$ and $\pi_1,\ldots,\pi_p$ are the dimensionless group. Buckingham's \textpi\ theorem brings a problem to its fundamental form through the \textpi\ groups \cite{Curtis}, reducing the quantities involved \cite{Fatoyinbo} and decreasing the complexity of the mathematical analysis of the problem of interest.

In the scientific literature, there are many original applications where Buckingham’s \textpi \ theorem has been successfully applied. Despite being proposed a long time ago, this theorem is still applied in science and engineering (\lq\lq ... old but not obsolete ...\rq\rq, \ paraphrasing the sci-fi movie Terminator Genisys) \cite{Terminator}. In \cite{Fatoyinbo}, dimensional analysis was applied to processing the biological cells by means of microfluidic devices. In \cite{Reddy}, using dimensionless groups, the authors studied the characteristics of different bearing parameters as temperature varies. Furthermore, in \cite{Ciulla} \textpi \ groups were adopted for a most effective description of the characteristic parameters of the thermal balance for the energy demand evaluation of a high-performance non-residential building. The creation of a rapid impedance model for proton exchange membrane fuel cells using physical and geometric parameters was analyzed in \cite{Russo}. The latter suggests defining dimensionless groups according to Buckingham's \textpi\ theorem so that the relationships between the fundamental dimensions and the physical variables involved in the process under discussion can be adequately described. This strategy was helpful in solving issues where first-principles models are unknown, challenging to build, or impossible to compute. In \cite{Hu}, methods were developed based on the use of Buckingham’s \textpi\ theorem for optimizing tests inside wind tunnels. In \cite{Bobbili} a comparative study on the  wire electrical discharge machining of reinforcement materials was realized by means of Buckingham's \textpi \ theorem. The theorem was used to model the input variables and thermophysical characteristics of wire electric discharge machining on the material removal rate and surface roughness of aluminum and steel.

Although this approach has been widely adopted in various physical applications, to the best of our knowledge this methodology has never been applied in the NDT\&E context. In this paper, we propose a new methodology to simultaneously estimate the thickness and the electrical conductivity of conductive plates by means of ECT. The proposed approach is based on dimensionless groups derived from the celebrated Buckingham’s \textpi \ theorem \cite{Buck1}.

This specific application is motivated by an acknowledgment that the measurement of the value thickness and electrical conductivity of conductive materials is a crucial factor in all production and manufacturing processes (e.g., heat treatment, rolling and pressing). Indeed, these two quantities directly affect the quality properties of finished products, such as hardness, toughness, and tensile strength \cite{Normativa,Conductivity1,Conductivity2, Conductivity3}. In this scenario, accurate and real-time monitoring of the thickness and electrical conductivity of conductive materials are essential to improve production quality and efficiency. In-line measurement techniques are essential because they enable automatic quality control during the production phase
%thus ensuring products or materials with appropriate levels of accuracy,
at reasonable prices and short inspection times, as required in the Industry 4.0 paradigm.

The possibility of applying ECT methods to the simultaneous estimation of several parameters, such as thickness, electrical conductivity and lift-off, has been widely studied in the literature. ECT methods are characterized by low-cost hardware and experimental set-up, contactless measurements, and insensitivity to non-conductive materials such as paints, dust, etc. In \cite{Phase_double_est} a method was proposed to simultaneously measure the thickness and electrical conductivity of the conductive sample, based on a single-frequency ECT method analyzing the phase of mutual impedance. ECT sensing systems using anisotropic magnetoresistive sensors for simultaneous estimation of thickness and electrical conductivity was proposed in \cite{MFD_double_est1}, while in \cite{MFD_double_est2} a new eddy current sensing method with a material-independent model for coupled-parameter estimation was proposed. An improved Newton iterative method to detect thickness, electrical conductivity, permeability, and lift-off of the conductive sample based on multi-frequency excitation ECT was developed in \cite{Multi_param_est}. Pulsed Eddy Current (PEC) techniques were also analyzed for multi-parametric estimation. For instance, in \cite{PEC_double_est1} the possibility to determine the thickness and electrical conductivity of conductive coatings on conductive plates was investigated using a PEC method, while in \cite{PEC_double_est2} a transient eddy-current measurement approach was proposed. 

In this paper, we use Buckingham’s \textpi \ theorem to simultaneously estimate the thickness and electrical conductivity of conductive plates. In particular, the major contributions of the paper are:

\begin{itemize}
    \item a relationship in term of dimensionless groups expressing the measured quantity as a function of the physical variables affecting it, in a frequency domain ECT experiment. We assume that the measured quantity is the self or mutual impedance of the probe. This assumption does not restrict the generality of the method;
%    \item a relationship in term of dimensionless groups between the measured quantity and the physical variables affecting it, in a frequency domain ECT experiment. We assume that the measured quantity is the self or mutual impedance of the probe. This assumption does not restrict the generality of the method;
    \item a method and its algorithm counterpart for the simultaneous estimation of the thickness and electrical conductivity by using dimensionless groups;
    \item a versatile experimental set--up for the simultaneous estimation of the thickness and electrical conductivity, based on single-frequency or multi-frequency measurements.
\end{itemize}

Compared with methods already established in the literature, the methodology proposed in this paper offers a number of advantages. First, Buckingham’s \textpi \ theorem makes it possible to reduce the number of variables to be considered, thus leading to a reduction in the problem's computational complexity. Second, it makes it possible to establish the structure of the relationships existing between the variables involved. Third, the proposed procedure is compliant with applications where the simultaneous estimation is required under in-line and real-time industrial conditions. Fourth, the proposed approach guarantees excellent accuracy. 

The paper is organized as follows. In Section \ref{section:theorical_background} we briefly summarize Buckingham’s \textpi \ theorem and show an example of its application in modelling a simple RLC circuit. In Section \ref{section:pi_ect} we apply Buckingham’s \textpi \ theorem to the specific problem and derive the essential structure of the relationship between the relevant variables in terms of \textpi \ groups. Section \ref{section:dimensionless_analysis} provides the method for the simultaneous estimation of the thickness and electrical conductivity of a nonmagnetic plate. Section \ref{section:experimental_charact} contains descriptions of the experimental set--up, case studies and experimental results. Finally, some conclusions are drawn in Section \ref{section:Conclusion}.

\section{Notations} \label{section:notations}
In this work we assume the following standard notations:
\begin{description}
\item[$\Bar{X}$] represents a complex number;
\item[$\Dot{Z}$] (upper-case) represents a complex impedance value;
%\item[$\Re \{ \cdot \}$] is the real part operator;
%\item[$\Im \{ \cdot \}$] is the imaginary part operator;
\item[$\mathbf{v}$] represents a real valued vector.
\end{description}

\section{Buckingham’s \textpi \ theorem} \label{section:theorical_background}
Dimensional analysis includes the set of all methods useful in reducing the dimension of the manifold representing a physical problem, before carrying out a quantitative analysis. Buckingham’s \textpi \ theorem (1914) is a fundamental \lq\lq tool\rq\rq \ to achieve this result \cite{Buck1}. Its root lies in previous publications by Lord Rayleigh (1877), J. Bertrand (1878), A. Vaschy (1892) and D. Riabouchinsky (1911). In essence, Buckingham’s \textpi \ theorem states that any physical law can be expressed in terms of dimensionless parameters, called \lq\lq \textpi \ groups\rq\rq \ , since physical laws are independent of the system of units.

Buckingham's \textpi \ theorem is stated as follows.

\begin{theorem}[Buckingham's \textpi \ theorem] \label{th:buckingham}
Let a physical problem involving $n$ dimensional scalar variables be modeled by a scalar equation of the type:
\begin{equation}
    g \left(q_1, q_2, q_3, \ldots ,q_n \right) = 0.
    \label{eq:buck2}
\end{equation}
Let the physical dimensions of all variables expressed in term of a set of $k$ fundamental dimensions $D_1,\ldots,D_k$:
\begin{equation}
    \dim q_i= D_1^{a_{i1}} \times \ldots \times D_k^{a_{ik}},\, i=1,\ldots,n,
\end{equation}
then there exist $p=n-k$ dimensionless groups \textpi$_1$, \textpi$_2$, $\ldots$ , \textpi$_p$ such that \eqref{eq:buck2} can be cast in the form
\begin{equation}
    G \left(\pi_1, \pi_2, \pi_3, \ldots, \pi_p \right) = 0.
    \label{eq:buck5}
\end{equation}
\end{theorem}

Theorem \ref{th:buckingham} can be easily extended to other cases where the laws of physics are described by vectors (tensors) quantities and/or multiple equations.

Buckingham's \textpi \ theorem does not give the explicit expression for $G$, given $g$, but, rather, has to be derived explicitly, after the \textpi \ groups have been computed.

An example of physical dimensions is given by those from the SI base units: \textsf{T} (time), \textsf{L} (length), \textsf{M} (mass), \textsf{I} (electric current), \textsf{$\Theta$} (absolute temperature), \textsf{N} (amount of substance) and \textsf{J} (luminous intensity). However, any set of fundamental dimensions can be used in Theorem \ref{th:buckingham}.

\begin{remark}
Buckingham's \textpi \ theorem for the special case when there is one dependent variable, i.e. $q_1=f\left( q_2,\ldots,q_n \right)$ gives
\begin{equation}
    \pi_1=F \left( \pi_2,\ldots,\pi_p \right).
\end{equation}
Similarly, Buckingham's \textpi \ theorem extends can be extended to the case of two or more dependent variables.
\end{remark}

A possible method to find the dimensionless groups is described in detail in Appendix \ref{app:construction}. The RLC series circuit provides an example of application to make the concepts crystal clear. In a RLC circuit the phasor $\Bar{I}$ of the electrical current circulating in the elements (see Figure \ref{fig:RLC_Circuit}) is a function of the phasor of the voltage generator and of the passive components, i.e.
\begin{equation}
\label{eq:rlc}
    \Bar{I}=f \left( \Bar{E},\omega,R,L,C\right)
\end{equation}
that, as shown in Appendix \ref{app:construction}, can be cast in dimensionless form as:
\begin{equation}
\label{eq:rlc_pi}
    \Bar{\pi}_1=F \left( \pi_2,\pi_3\right)
\end{equation}
where
\begin{equation}
\label{eq:pi_groups_RLC}
    \Bar{\pi}_1=\frac{R\Bar{I}}{\Bar{E}}, \ \pi_2=\frac{\omega L}{R}, \ \pi_3=\frac{1}{\omega R C}.
\end{equation}

\begin{figure}[htb]
    \centering
    \includegraphics[width=0.30\columnwidth]{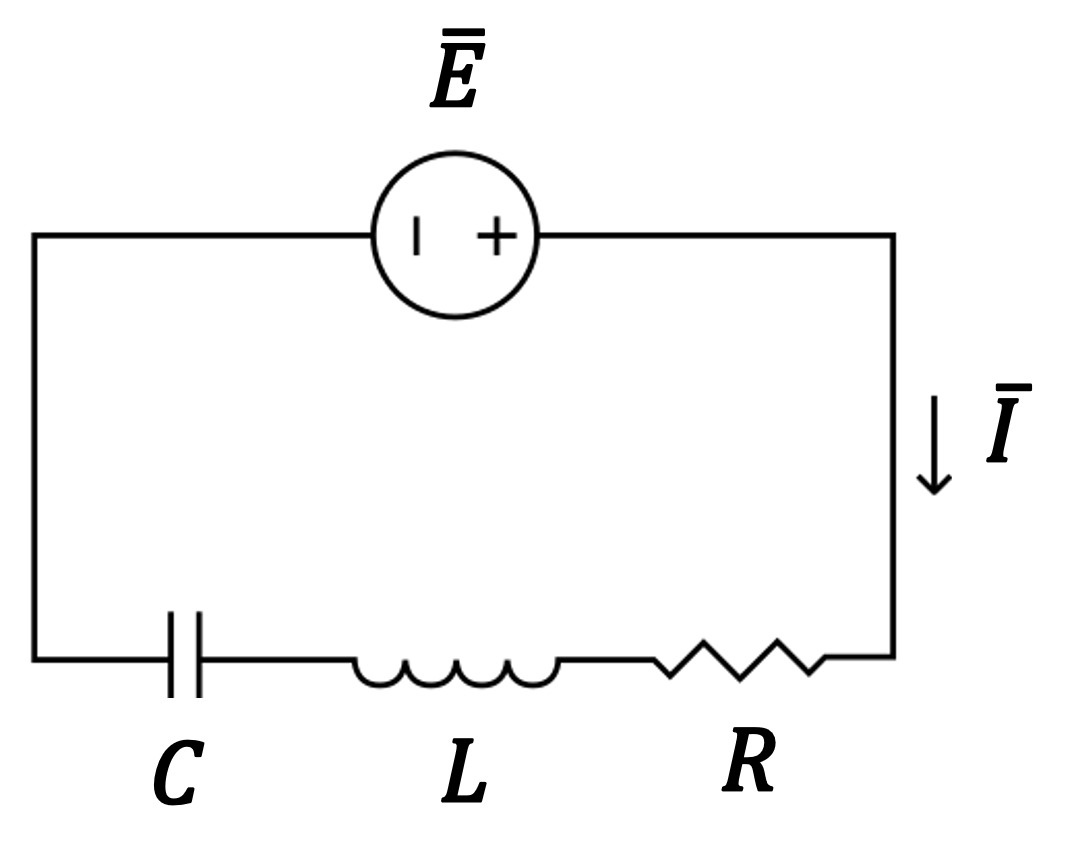}
    \caption{The RLC series circuit in time-harmonic operation. The angular frequency is $\omega$.}
    \label{fig:RLC_Circuit}
\end{figure}

The total number of physical variables required to describe the RLC series circuit reduces from $n=6$ in equation \eqref{eq:rlc} to $p=3$ in equation \eqref{eq:rlc_pi}, because the dimensions of $\Bar{I},\Bar{E},\omega,R,L,C$ can be expressed through a set of $k=3$ physical dimensions related to time, voltage and current.

The derivation of \eqref{eq:rlc_pi} and \eqref{eq:pi_groups_RLC} does not rely on explicit knowledge of $f$. However, Buckingham's \textpi \ theorem does not provide the explicit expression for $F$, but rather, it must be derived from knowledge of $f$ and the related \textpi \ groups. For the RLC circuit, from
\begin{equation}
    \Bar{I}=\frac{\Bar{E}}{R+j\omega L -j/\omega C}
\end{equation}
and the \textpi \ groups in \eqref{eq:pi_groups_RLC}, it can be easily found that
\begin{equation}
    \frac{R \Bar{I}}{\Bar{E}}=\frac{1}{1+j\frac{\omega L}{R} -j \frac{1}{\omega R C}},
\end{equation}
i.e. $F(\pi_2,\pi_3)=1/\left[1+ j \left( \pi_2 - \pi_3\right) \right]$.

\begin{remark}\label{rem:nuq}
It is worth noting that the \textpi \ groups are not unique. The same law of physics can be expressed by means of many different sets of \textpi \ groups. Indeed, products, powers and ratios of \textpi \ groups are still \textpi \ groups. The selection of the proper set of \textpi \ groups is related to the \emph{specific} application and the related solution method. This makes dimensional analysis via Buckingham's \textpi \ theorem extremely versatile. The specific procedure sketched in Appendix \ref{app:construction} is one of several possible equivalent choices. It guarantees that all the independent variables, apart from the repeating ones, appears in \emph{one and only one} \textpi \ group.
\end{remark}

The main advantage of Buckingham’s \textpi \ theorem consists in the reduction of the number of relevant variables describing a problem from $n$ to $p$, where $p=n-k$. This approach is especially effective when $n$ is of the order of a few units. This is because $k$ is of the order of few units, $k \le 7$ in the SI, thus making $p$ a fraction of $n$ when $n$ is of the order of few units. For instance, by analyzing the RLC circuit we have $n=6$ and $k=3$, thus yielding $p=3$, that is one-half of $n$. This has a major impact in reducing the amount of experimental and/or numerical data required to make correlations of physical variables. Indeed, in order to fully characterize the function $f$ appearing in \eqref{eq:rlc_pi}, the parameters array $\left( \Bar{E},\omega,R,L,C \right)$ has to be varied in $\mathbb{C} \times \mathbb{R}^4$, whereas the full characterization of the function $F$ in \eqref{eq:rlc_pi} requires the parameters array $\left( \pi_2,\pi_3 \right)$ to be varied in $\mathbb{R}^2$. This latter option (evaluation of $F$) is definitely less expensive, in terms of the number of experimental tests or numerical simulations, if compared to the former (evaluation of $f$). Summing up, dimensional analysis is an extremely powerful technique for formulating physics problems in their most basic forms by minimizing the degrees of freedom of the problem. 

Another advantage offered by Buckingham’s \textpi \ theorem is that its application does not require a priori knowledge of the law relating the key physical quantities. It can be applied starting from only the knowledge of the physical variables describing the phenomenon. This is very important when the laws constraining the physical quantities are unknown, and it provides a guide for finding such laws.

\section{Dimensional analysis in Eddy Current Testing} \label{section:pi_ect}
The possibility of applying ECT to the simultaneous estimation of multiple parameters, such as thickness and electrical conductivity, has long attracted many researchers' interest.

An Eddy Current Probe (ECP) typically comprises a driving coil that generates a time-varying magnetic flux density, which in turn induces a current density in a conductive material, and a receiver coil or field sensor to sense the reaction magnetic flux density due to the induced eddy currents. Probes of different shapes and arrangements may also be adopted, based on a single or multiple coils to produce the driving field and to measure the response of the material being tested \cite{Phase_double_est,Multi_param_est,PEC_double_est1,PEC_double_est2}: one coil for the driving field and a magnetic flux density sensor to sense the response \cite{MFD_double_est1,MFD_double_est2}, and so on. Tests can be performed by adopting single-frequency or multiple-frequency approaches and analyzing the various quantities measured, such as the variation in magnetic flux density due to the presence of the material being tested or the variation in self-impedance (for a single coil application) or mutual-impedance (for a multiple coil application). In all these ECT scenarios, Buckingham’s \textpi \ theorem can be profitably applied for a thorough understanding of the relationships between the variables describing the physical problem and to simplify the model by reducing the number of degrees of freedom required to describe the system.

The effectiveness of Buckingham’s \textpi \ theorem in the simultaneous evaluation of thickness and conductivity via ECT data can be demonstrated by a specific case study without loss of generality. This involves an ECP comprising two coaxial coils (T/R configuration) with the upper coil used as the driving coil and the lower coil as the pick-up coil. Coaxial coils and nonmagnetic materials have been assumed. Figure \ref{fig:Probe} shows the geometry of the problem, with both the conductive plate and the ECP.

In the case of interest (T/R coil configuration), the measured quantity is $\Delta\dot{Z}_{m} = \dot{Z}_{m,plate}-\dot{Z}_{m,air}$, i.e. the difference in mutual impedance between the coils when the ECP is located on the plate ($\dot{Z}_{m,plate}$) and in air ($\dot{Z}_{m,air}$), respectively, at a prescribed angular frequency. Hereafter we assume coaxial coils and nonmagnetic materials. The key physical quantities determining $\Delta \dot{Z}_{m} ( \omega )$ are (see Figure \ref{fig:Probe}):
\begin{itemize}
    \item the parameters describing the geometry of the probe: the internal $r_1$ and external $r_2$ radii of the coils, the height $h_1$ and number of turns $N_1$ of the receiving coil, the height $h_2$ and number of turns $N_2$ of the driving coil and the separation between the coils $d$;
    \item the angular frequency $\omega$ of the driving current applied for the test;
    \item the thickness $\Delta h$ and the electrical conductivity $\sigma$ of the conductive plate;
    \item the magnetic permeability of the vacuum $\mu_0$ and the corresponding magnetic reluctance $\nu_0=1/\mu_0$.
    \item the lift-off $l_0$ between the plate and the ECP and the tilting of the ECP probe w.r.t. the perpendicular to the plate.
\end{itemize}

The geometrical parameters of the ECP are normalized with respect to a length $D=r_2$ and are grouped in a dimensionless vector $\mathbf{t}=\left( r_1/D, h_1/D, h_2/D, d/D \right)$. The normalization constant $D$ represents the size of the probe. Other choices for $D$ can equally be made.

\begin{figure}[htb]
    \centering
    \includegraphics[width=0.95\columnwidth]{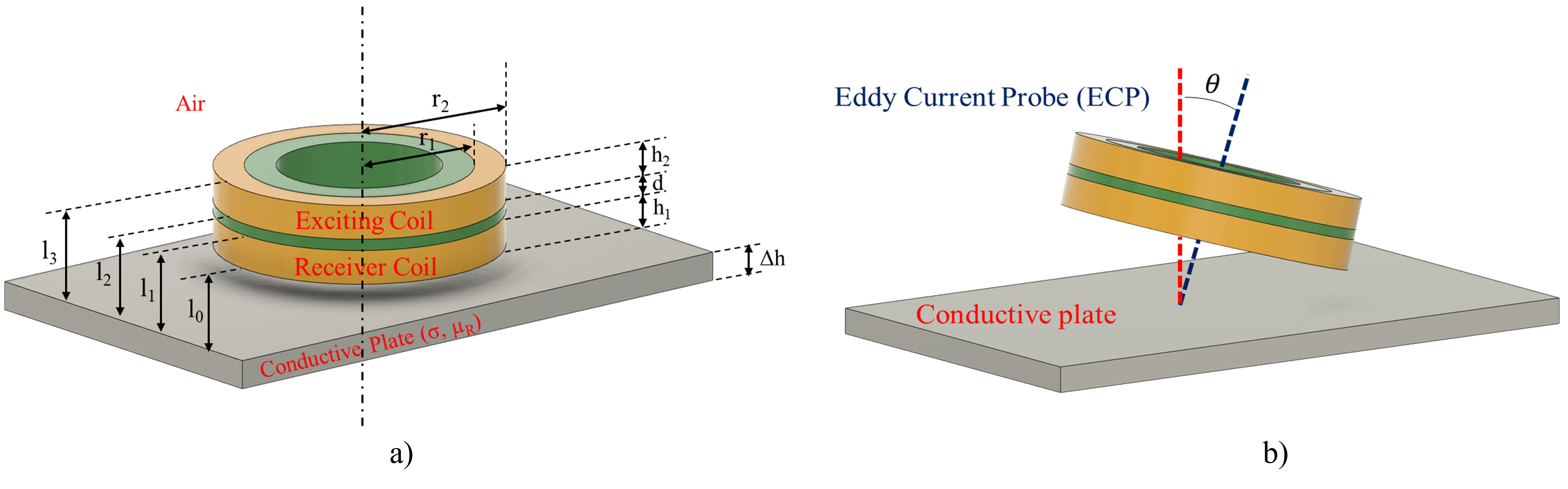}
    \caption{Representation of the axis-symmetrical ECP placed on the conductive plate and their geometrical characteristics.}
    \label{fig:Probe}
\end{figure}

All the listed parameters (probe geometry, number of turns in the coils, conductivity and thickness of the plate, lift-off and tilting of the ECP with respect to the plate) affect the mutual impedance between the transmitting and receiving coils, i.e. 
\begin{equation} 
    \frac{\Delta \dot{Z}_{m}}{N_1 N_2} = f \left( \omega,\sigma,\nu_0,\Delta h,D,\textbf{t},l_0,\theta \right).
    \label{eq:Z_relation}
\end{equation}

The evaluation of $f$, either by a numerical method or an experimental campaign, is no simple task, because its cost/time increases exponentially with the number of arguments.

Equation \eqref{eq:Z_relation} involves a total of nine variables ($n=9$), with seven real-valued and scalar independent variables $\omega,\sigma,\Delta h, D, l_0, \theta,\nu_0$, one real-valued vector independent variable $\mathbf{t}$ and one complex-valued dependent variable $\Delta \dot{Z}_{m}/N_1 N_2$. Variables $D$ and $\mathbf{t}$ correspond to five real-valued and scalar variables $(r_1,r_2,h_1,h_2,d)$. All variables can be expressed in terms of three fundamental dimensions ($k=3$), length $L$, time $T$ and impedance $\Omega$, as shown in Table \ref{tab:Variable_thickness}.

\begin{table}[htb]
\centering
\captionsetup{justification=centering}
\caption{Dimensional variables in the multi-parameter estimation problem, expressed in terms of fundamental dimensions.}
\begin{tabular}{cccc}
\hline 
\hline
Numbers & Parameters                          & Symbols              & Fundamental dimensions         \\ \hline
1       & impedance variation                 & $\Delta \dot{Z}_{m}$ & {[}$L^0 T^0 \Omega^1${]}       \\
2       & electric conductivity               & $\sigma$             & {[}$L^{-1} T^0 \Omega^{-1}${]} \\
3       & magnetic reluctance of the   vacuum & $\nu_0$         & {[}$L^1 T^{-1} \Omega^{-1}${]} \\
4       & angular frequency                   & $\omega$             & {[}$L^0 T^{-1} \Omega^0${]}    \\
5       & thickness                           & $\Delta h$           & {[}$L^1 T^0 \Omega^0${]}       \\
6       & lift-off                            & $l_0$                & {[}$L^1 T^0 \Omega^0${]}       \\
7       & ECP characteristic length               & D                    & {[}$L^1 T^0 \Omega^0${]}       \\
8       & ECP shape          & \textbf{t}           & {[}$L^0 T^0 \Omega^0${]}       \\
9       & probe orientation features          & $\theta$    & {[}$L^0 T^0 \Omega^0${]}       \\
\hline \hline
\label{tab:Variable_thickness}
\end{tabular}
\end{table}

Buckingham’s \textpi \ theorem makes it possible to obtain six dimensionless groups ($p=n-k=6$) as listed in Table \ref{tab:pi_groups_thick_final} (see Appendix \ref{app:construction} for details), where $\nu_0$, $\omega$ and $D$ have been assumed as repeating variables. Each dimensionless variable of the original problem, $\mathbf{t}$ and $\theta$, is assigned to a dimensionless \textpi \ group, i.e. $\mathbf{\pi}_5=\mathbf{t}$ and $\pi_6=\theta$. Consequently, \eqref{eq:Z_relation} can be expressed as $    \pi_1 = F \left( \pi_2,\pi_3,\pi_4,\pi_5,\pi_6 \right)$, that is
\begin{align}
    \frac{\Delta \dot{Z}_{m} \nu_0}{N_1 N_2 \omega D} & = F \left( D \sqrt{\frac{\omega \sigma} {2 \nu_0}},\frac{\Delta h}{D},\frac{l_0}{D},\mathbf{t},\theta \right) \label{eq:Thick3}\\ 
     & = F \left( \frac{D}{\delta},\frac{\Delta h}{D},\frac{l_0}{D},\mathbf{t},\theta \right), \label{eq:Thick2}
\end{align}
where $F$ is a proper function and the skin-depth $\delta$ is equal to
\begin{equation}
    \delta = \sqrt{\frac{2 \nu_0}{\omega \sigma}}.
    \label{eq:depht_pen}
\end{equation}
The choice of the \textpi \ groups made in Table \ref{tab:pi_groups_thick_final} is not unique, as highlighted in Remark \ref{rem:nuq}. The choice of a specific set of \textpi \ groups depends on the specific application.

\begin{table}[htb]
\centering
\captionsetup{justification=centering}
\caption{List of the \textpi \ groups for the case of interest.}
\begin{tabular}{lll}
\hline \hline
\multicolumn{3}{c}{Dimensionless groups} \\
\hline
$\Bar{\pi}_1 = \frac{\Delta \dot{Z}_m \nu_0} {N_1 N_2 \omega D}$ & $\pi_2 = D \sqrt{\frac{\omega \sigma} {2 \nu_0}}$ & 
$\pi_3 = \frac{\Delta h}{D}$ \\
$\pi_4 = \frac{l_0}{D}$ & $\pi_5= \mathbf{t}$ & $\pi_6 = \theta$ \\
\hline \hline
\label{tab:pi_groups_thick_final}
\end{tabular}
\end{table}

By taking into account that the purpose of this study is to measure the thickness and the electrical conductivity of a conductive plate, given the characteristics $\mathbf{t}$ of the ECP and the lift-off $l_0$ and tilting $\theta$, it is convenient to focus on groups $\Bar{\pi}_1$, $\pi_2$ and $\pi_3$ since $\pi_4$, $\pi_5 $ and $\pi_6$ are known. Consequently, the final dimensionless relationship under analysis is:
\begin{equation}
    \frac{\Delta \dot{Z}_{m} \nu_0}{N_1 N_2 \omega D} = F \left( D \sqrt{\frac{\omega \sigma} {2 \nu_0}},\frac{\Delta h}{D} \right),
    \label{eq:pi_final}
\end{equation}
where, with a slight abuse of notations, the values of $\pi_4$, $\pi_5 $ and $\pi_6$ are understood.

Equation \eqref{eq:pi_final} has to be compared with the counterpart of \eqref{eq:Z_relation} for prescribed (understood) $\mathbf{t}$, $l_0$ and $\theta$:
\begin{equation} 
    \frac{\Delta \dot{Z}_{m}}{N_1 N_2} = f \left( \omega,\sigma,\nu_0,\Delta h,D \right).
    \label{eq:f_reduced}
\end{equation}

The impact of Buckingham's \textpi \ Theorem is relevant. First, by starting from \eqref{eq:f_reduced}, which involves a complex function of five real arguments, it is possible to obtain an equivalent relationship requiring a complex function $F$ of two real arguments, without the explicit knowledge of the original function $f$. The new (reduced) function $F$ can be easily computed numerically or measured experimentally, since it is defined in $\mathbb{R}^2$ rather than in $\mathbb{R}^5$. Moreover, by combining \eqref{eq:pi_final}, with the \textpi \ groups of Table \ref{tab:pi_groups_thick_final}, it is easy to realize that $f$ and $F$ are closely related. Indeed, it results that
\begin{equation}
    f \left( \omega,\sigma,\nu_0,\Delta h,D \right) = \frac{\omega D}{\nu_0} F \left( D \sqrt{\frac{\omega \sigma} {2 \nu_0}},\frac{\Delta h}{D} \right),
\end{equation}
thus the computation or experimental evaluation of $F$ gives the values of $f$.

Second, Buckingham's \textpi \ Theorem allows the inverse problem to be represented in the two-dimensional $(\pi_2,\pi_3)$ plane, rather than in a five-dimensional space. The latter remark underpins the method for the simultaneous estimation of $\sigma$ and $\Delta$ via level curves, described in Section \ref{section:dimensionless_analysis}.

Finally, the same approach can be applied to the case of a single coil ECP or to the case in which the reaction magnetic flux density is measured by a field sensor.

\section{Simultaneous estimate of thickness and electrical conductivity via dimensionless groups} \label{section:dimensionless_analysis}
In this Section the effectiveness of the concept of dimensional analysis is demonstrated with reference to the problem of the simultaneous estimation of the electrical conductivity and thickness of a metallic plate. This is a relevant problem, from the practical perspectives.

This Section is organized in two parts. The first part proposes an approach based on level curves, which can be easily introduced, thanks to the key role played by the \textpi \ groups. The second part is devoted to \lq\lq translating\rq\rq \ the method's physical limits in terms of level curves.

Without loss of generality, a planar geometry has been considered in order to demonstrate the effectiveness of dimensional analysis. The same treatment can be applied to non-planar geometries like tubes.

\subsection{\textpi \ groups, level curves and estimation method}
Thanks to the abstract representation of \eqref{eq:pi_final}, where the (complex) \emph{measured quantity} $\Bar{\pi}_1$ is a function defined in the $(\pi_2,\pi_3)$ plane, it is possible to introduce a set of level curves with respect to $\Bar{\pi}_1$. This is possible because the measured quantity $\dot{Z}_{m}$ and the unknowns $\sigma$ and $\Delta h$ are not mixed in the \textpi \ groups. 

In order to obtain a level curve in the $(\pi_2,\pi_3)$ plane, it is necessary to prescribe the value of a real valued quantity. For instance, the real part, the imaginary part, the magnitude, the phase of $\Bar{\pi}_1$, or any other real function of $\Bar{\pi}_1$ can be chosen. Figure \ref{fig:Trend_adimensional} shows the level curves for the four basic quantities: $\Re{\Bar{\pi}_1}$, $\Im{\Bar{\pi}_1}$, $|\Bar{\pi}_1|$ and $\phase{\Bar{\pi}_1}$, in the $(\pi_2,\pi_3)$ plane. The parameters for the underlying ECP are provided in Table \ref{tab:geometrical_param}.

\begin{figure}[htb]
    \centering
    \includegraphics[width=0.95\columnwidth]{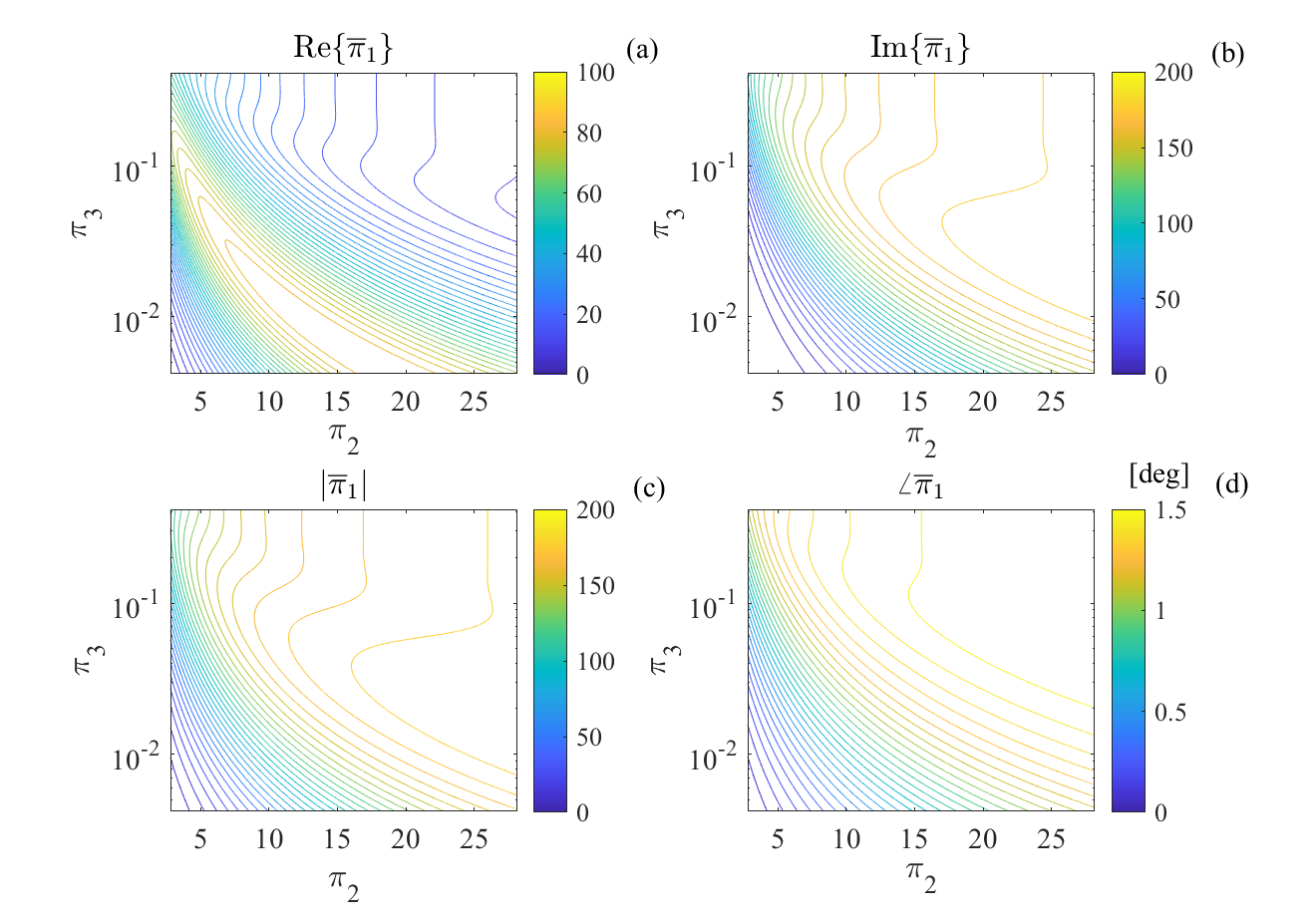}
    \caption{Level curves for $\Bar{\pi}_1$: (a) $\Re{\Bar{\pi}_1}$, (b) $\Im{\Bar{\pi}_1}$, (c) $|\Bar{\pi}_1|$, and (d) $\phase{\Bar{\pi}_1}$. The level curves were plotted at constant step, i.e. the difference between the values for two consecutive level curves is constant. Smaller gradients are found where the distance between the curves increases.}
    \label{fig:Trend_adimensional}
\end{figure}

\begin{table}[htb]
\centering
\captionsetup{justification=centering}
\caption{Values of the parameters of the ECP.}
\begin{tabular}{cc}
\hline \hline
Parameter  & Value  \\ \hline
$h_1 = h_2$ & 6   {[}mm{]}           \\
$d$         & 2.20  {[}mm{]}           \\
$r_1$       & 23.60 {[}mm{]}            \\
$r_2$       & 23.95 {[}mm{]}            \\
$l_0$       & 1 {[}mm{]}             \\
$N_1 = N_2$ & 17             \\ 
$\theta$ & 0 {[}°{]}             \\ \hline \hline
\label{tab:geometrical_param}
\end{tabular}
\end{table}

The level curves in Figure \ref{fig:Trend_adimensional} were obtained from a numerical evaluation of $F(\pi_2,\pi_3)$ in a range of values for $\pi_2$ and $\pi_3$. The numerical evaluation was carried out by evaluating $\Delta \dot{Z}_{m}$ via the the semi-analytical model by Dodd and Deeds \cite{Dodd_deeds}. $\pi_2$ was varied in the range $\left[2.82;\ 28.2\right]$, whereas $\pi_3$ in the range $\left[4.2 \times 10^{-3}; \    42 \times 10^{-3}\right]$. These ranges were obtained by retaining $\sigma$ constant and varying $\Delta h$ and $\omega$, as in Table \ref{tab:physical_param}. The thickness $\Delta h$ was varied with increments of 10 $\mu$m, whereas the frequency was varied with increments of 2 Hz.

\begin{table}[htb]
\centering
\captionsetup{justification=centering}
\caption{The physical parameters used for the numerical evaluation of $F(\cdot,\cdot)$.}
\begin{tabular}{cc}
\hline \hline
Parameters & Value                    \\ \hline
$\sigma$    & 35 MS/m          \\
$\Delta h$ & 0.1 ÷ 10 {[}mm{]}  \\
$\omega/2\pi$        & 100 ÷  10000 {[}Hz{]} \\ \hline \hline
\label{tab:physical_param}
\end{tabular}
\end{table}

Level curves are a powerful tool to solve equation
\begin{equation}
\label{eq:main_eq}
    F(\pi_2,\pi_3) = \Bar{\pi}_1.
\end{equation}
Specifically, from the measurement of $\dot{Z}_{m}$ at a prescribed angular frequency $\omega$, it is possible to compute $\Bar{\pi}_1$ as $\Delta \dot{Z}_{m} \nu_0 /N_1 N_2 \omega D$. Then, from the specific value of $\Bar{\pi}_1$, it is possible to solve \eqref{eq:main_eq} by finding the intersection point between the level curves from two or more different plots of Figure \ref{fig:Trend_adimensional}, as shown in Figure \ref{fig:Intersection_single}.
\begin{figure}[htb]
    \centering
    \includegraphics[width=0.7\columnwidth]{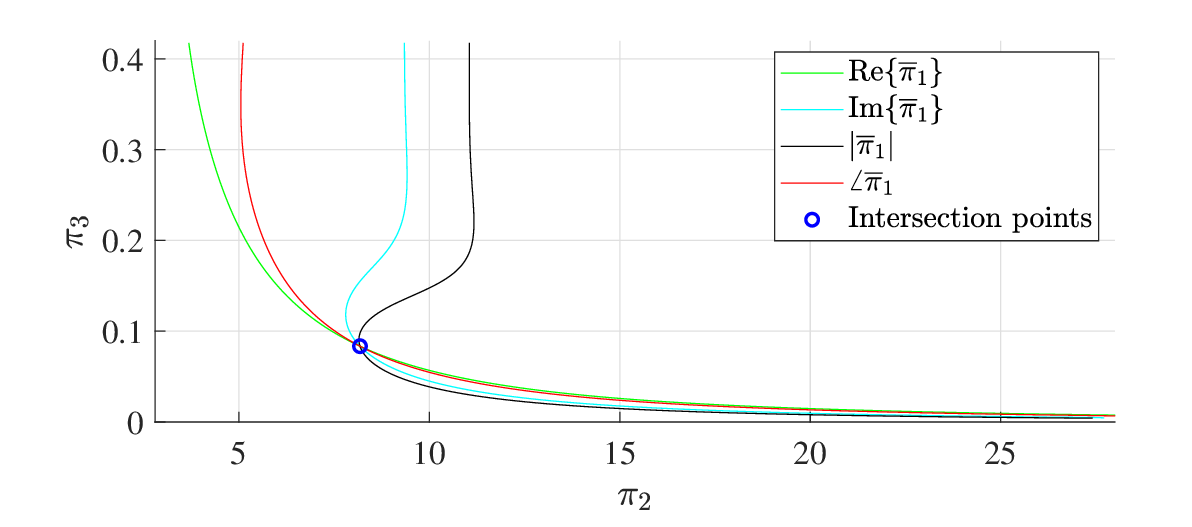}
    \caption{Intersection of level curves from the mutual impedance measured at an individual prescribed frequency. The plate has an electrical conductivity of $18$ MS/m and a thickness of $2$ mm. The frequency is 1650 Hz.}
    \label{fig:Intersection_single}
\end{figure}
Once $\pi_2$ and $\pi_3$ have been evaluated, the unknown electrical conductivity $\sigma$ and thickness $\Delta h$ can be evaluated as
\begin{align}
    \sigma  & =\frac{2 \nu_0}{\omega} \left( \frac{\pi_2}{D} \right)^2  \label{eq:sigma} \\
\Delta h & = D \pi_3. \label{eq:dh}
\end{align}

The step-by-step algorithm is:
\begin{itemize}
    \item Measure $\Delta \dot{Z}_{m}$ at a prescribed $\omega$;
    \item compute $\Bar{\pi}_1=\Delta \dot{Z}_{m} \nu_0 /N_1 N_2 \omega D$;
    \item find the level curves for at least two plots of Figure \ref{fig:Trend_adimensional};
    \item find the intersection point $(\pi_2,\pi_3)$ for the selected level curves;
    \item compute $\sigma$ and $\Delta h$ via \eqref{eq:sigma} and \eqref{eq:dh}.
\end{itemize}

Summing up, dimensional analysis makes it possible to cast the problem of retrieving $\sigma$ and $\Delta h$ in very simple terms as the intersection of level curves in a plane. This is because the five primary parameters $(\omega,\sigma,\nu_0,\Delta h, D)$ influencing the measured data combine in the very compact form given by groups $\pi_2$ and $\pi_3$, rather than individually as in \eqref{eq:f_reduced}. For the same reason, i.e. that the influence parameters combine in a compact form, it is computationally feasible to numerically compute the function $F(\cdot,\cdot)$, which is a function of two parameters rather than $f(\cdot,\cdot,\cdot,\cdot,\cdot)$, which depends on five parameters. Finally, $F(\cdot,\cdot)$ can be pre-computed and stored once for all, given $\mathbf{t}$, $l_0$ and $\theta$.

\subsection{Processing multiple measurements}
Noise is a major issue when dealing with experimental data. Method accuracy can be increased by processing the impedance measured at multiple frequencies. In this case, the level curves intersection procedure has to be repeated at each angular frequency, thus obtaining a set of points in the $(\pi_2,\pi_3)$ plane, as shown in Figure \ref{fig:Intersection_trend}.

\begin{figure}[htb]
    \centering
    \includegraphics[width=0.7\columnwidth]{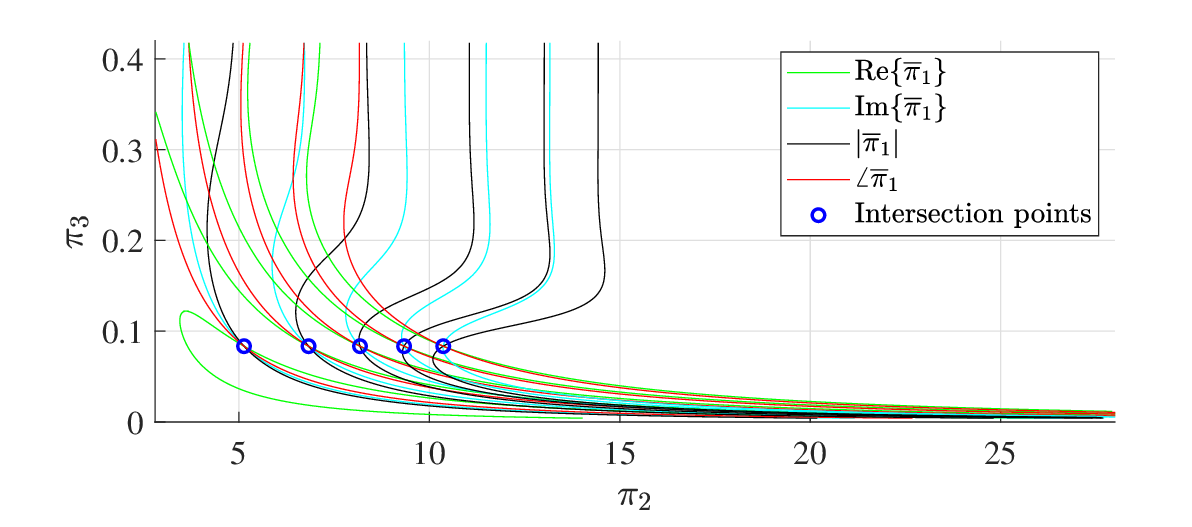}
    \caption{Intersection of level curves from the mutual impedance measured at five different frequencies. The plate has an electrical conductivity of $18$ MS/m and a thickness of $2$ mm. The frequencies are 650, 1150, 1650, 2150 and 2650 Hz.}
    \label{fig:Intersection_trend}
\end{figure}

Since $\pi_2$ is independent of $\omega$, whereas $\pi_3$ depends on $\omega$, as it is proportional to $\sqrt{\omega}$, the intersection points are distributed along a horizontal line. Each intersection point related to the impedance measured at the $i-$th angular frequency $\omega_i$ corresponds to an estimate $\sigma_i$ and $\Delta h_i$ of the electrical conductivity and thickness of the plate. The \lq\lq final\rq\rq \ estimate of $\sigma$ and $\Delta h$ can be obtained by processing all the $\sigma_i$s and the $\Delta h_i$s with an improved robustness as it  combines the information from different frequencies.

Alternatively, it is possible to plot the level curves on the $(\sigma,\Delta h)$ plane, by means of \eqref{eq:sigma} and \eqref{eq:dh}. The latter strategy is extremely convenient because \emph{all} the level curves intersect at the \emph{same} point regardless the angular frequency, as shown in Figure \ref{fig:Intersection_trend_norm}. The intersection point gives the estimate of $\sigma$ and $\Delta h$.
\begin{figure}[htb]
    \centering
    \includegraphics[width=0.7\columnwidth]{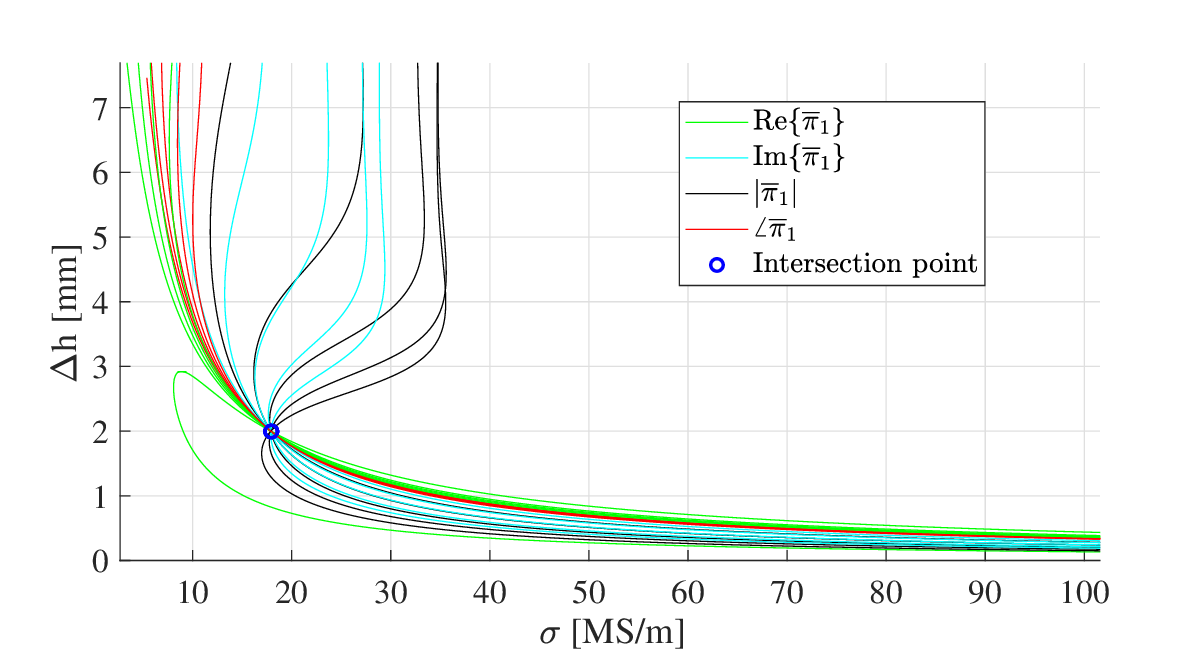}
    \caption{Representation on the normalized plane of the  frequency measurements obtained in the case of a plate with thickness of 2 mm and electrical conductivity of 18 MS/m  for the level curves  $\Re{\Bar{\pi}_1}$, $\Im{\Bar{\pi}_1}$, $|\Bar{\pi}_1|$, and $\phase{\Bar{\pi}_1}$ using five different excitation frequencies.}
    \label{fig:Intersection_trend_norm}
\end{figure}

\subsection{Regions of operation} \label{subsection:operating_area}
The $(\pi_2,\pi_3)$ plane can be divided into different regions, yielding different information that can be inferred from the measured data.

Three basic conditions must be considered:
\begin{enumerate}
    \item the skin-depth $\delta(\omega)$ is sufficiently smaller than $\Delta h$. i.e. $\pi_2 \pi_3$ is sufficiently larger than 1;
    \item the size of the probe $D$ is much smaller than $\Delta h$, i.e. $\pi_3 \gg 1$;
    \item the skin-depth $\delta(\omega)$ is much larger than $\Delta h$, i.e. $\pi_2 \pi_3 \ll 1$ and the size of the probe $D$ is much larger than $\Delta h$, i.e. $\pi_3 \ll 1$.
\end{enumerate}

In the first case (regions (c), (f) and (i) of Figure \ref{fig:Feasibility_ext}), the skin-depth is smaller than $\Delta h$ and this prevents the thickness $\Delta h$ being retrieved from the data (see \cite{Pinotti}), i.e. the dimensionless impedance $\Bar{\pi}_1$. This behaviour can be easily recognized in the plots of Figure \ref{fig:Trend_adimensional}, where the level curves become almost vertical, meaning that the $\pi_3$ does not affect the measured data $\Bar{\pi_1}$, i.e. $\Delta h$ cannot be retrieved from knowledge of $\Bar{\pi_1}$. We found numerically that the region where the level curves are almost vertical corresponds to $\pi_2 \pi_3 > 3$, as shown in Figure \ref{fig:Feasibility}. However, in these regions, the electrical conductivity can still be retrieved from $\Bar{\pi}_1$, since a change in the electrical conductivity, i.e. in $\pi_2$, determines a change in $\Bar{\pi}_1$.
\begin{figure}[htb]
    \centering
    \includegraphics[width=1\columnwidth]{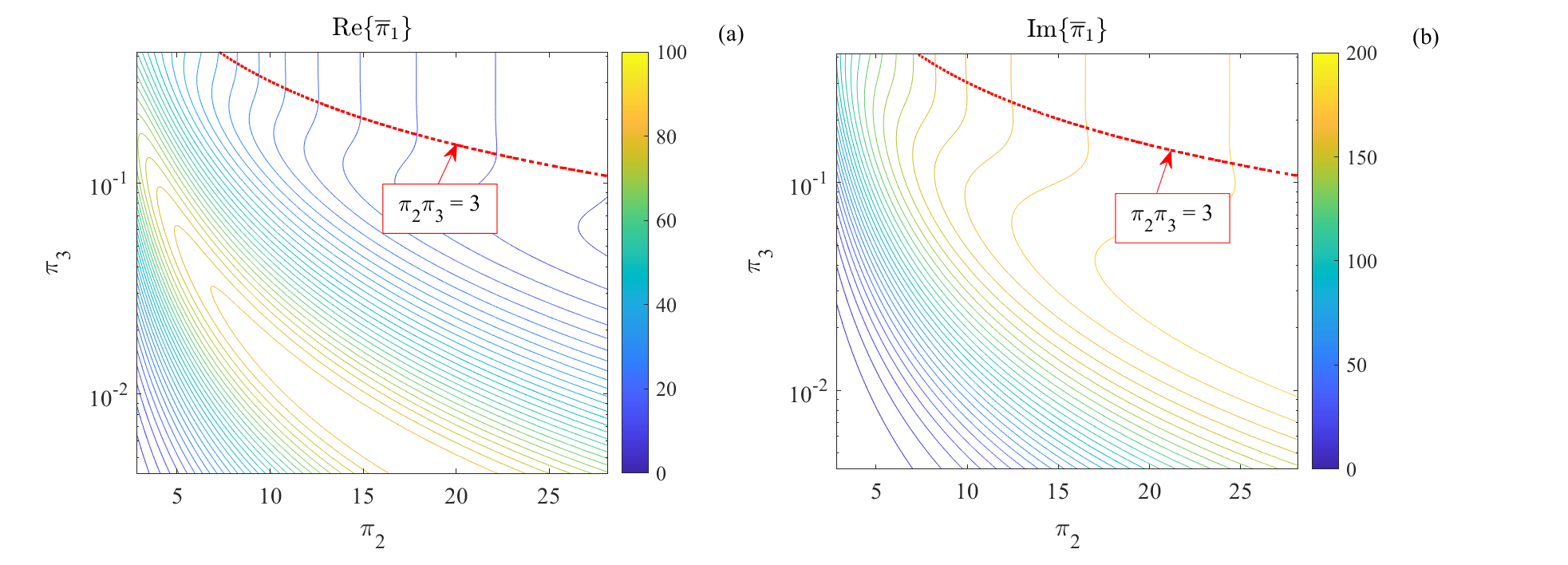}
    \caption{The level curves above the curve $\pi_2 \pi_3 = 3$ are almost vertical.}
    \label{fig:Feasibility}
\end{figure}

In the second case (regions (a), (b) and (c) of Figure \ref{fig:Feasibility_ext}), the probe is geometrically too small to interact with the bottom of the plate, regardless of the skin-depth, i.e. regardless of the value of $\pi_2=D/\delta$. This prevents the thickness $\Delta h$ being retrieved from $\Bar{\pi}_1$, but the electrical conductivity $\sigma$ can still be retrieved. As in the previous case, the level curves are almost vertical in regions (a), (b) and (c).

In the third case (region (g) of Figure \ref{fig:Feasibility_ext}, the probe is much larger than the thickness $\Delta h$ and the plate is fully penetrated by the electromagnetic field. In this case it is only possible to retrieve the surface electrical conductivity of the plate, i.e. the $\sigma \Delta h$ product, because the electromagnetic field only depends on this quantity, as highlighted in \cite{Sardellitti1,8715765}.

In the remaining regions (d), (e) and (h) of Figure \ref{fig:Feasibility_ext}, it is possible to retrieve both the electrical conductivity $\sigma$ and the thickness $\Delta h$, starting from the dimensionless impedance $\Bar{\pi_1}$. This is the so-called \emph{feasibility} region, where the largest amount of information can be retrieved from the measured data. Region (h) deserves to be highlighted, because it opens to the possibility of measuring both the electrical conductivity and the thickness of thin and very thin plates.

The plot of Figure \ref{fig:Feasibility} makes it possible to evaluate \textit{(i)} whether or not a measurement at a specific frequency has to be discarded and \textit{(ii)} whether or not the specific probe is suitable. Indeed, once the dimensionless impedance han been measured, if the related level curve does not cross a prescribed region(s) of interest, for instance region (d), (e) or (h) for the $\sigma$-$\Delta h$ estimate, then the measurement has to be discarded. Similarly, if several/many curves do not cross the region(s) of interest, it may be the case that \emph{all} the frequencies are out of the proper range or the size of the probe is not suitable for the specific plate. In any case, the key point is that the data itself makes it possible to understand whether or not the measurement is suitable to be processed, despite the lack of knowledge of the unknown parameters.

The plots of the level curves provide further precious but less recognized information. Specifically, at the higher frequencies where the thicknesses cannot be retrieved (regions (c), (f) and (i) of Figure \ref{fig:Feasibility_ext}), the spacing between the level curves increases, as shown in Figure \ref{fig:Trend_adimensional}. This means that the gradient of the measured data $\Bar{\pi}_1$ with respect to $\pi_2$ decreases for increasing $\pi_2$. In other words, whith large angular frequencies and/or electrical conductivities, the sensitivity of the measurement with respect to $\sigma$ decreases, even though some authors claimed that it is convenient to estimate the electrical conductivity in such conditions, because the data does not depend on the thickness $\Delta h$ \cite{CL_ECT_conductivity2}.

\begin{figure}[htb]
    \centering
    \includegraphics[width=0.80\columnwidth]{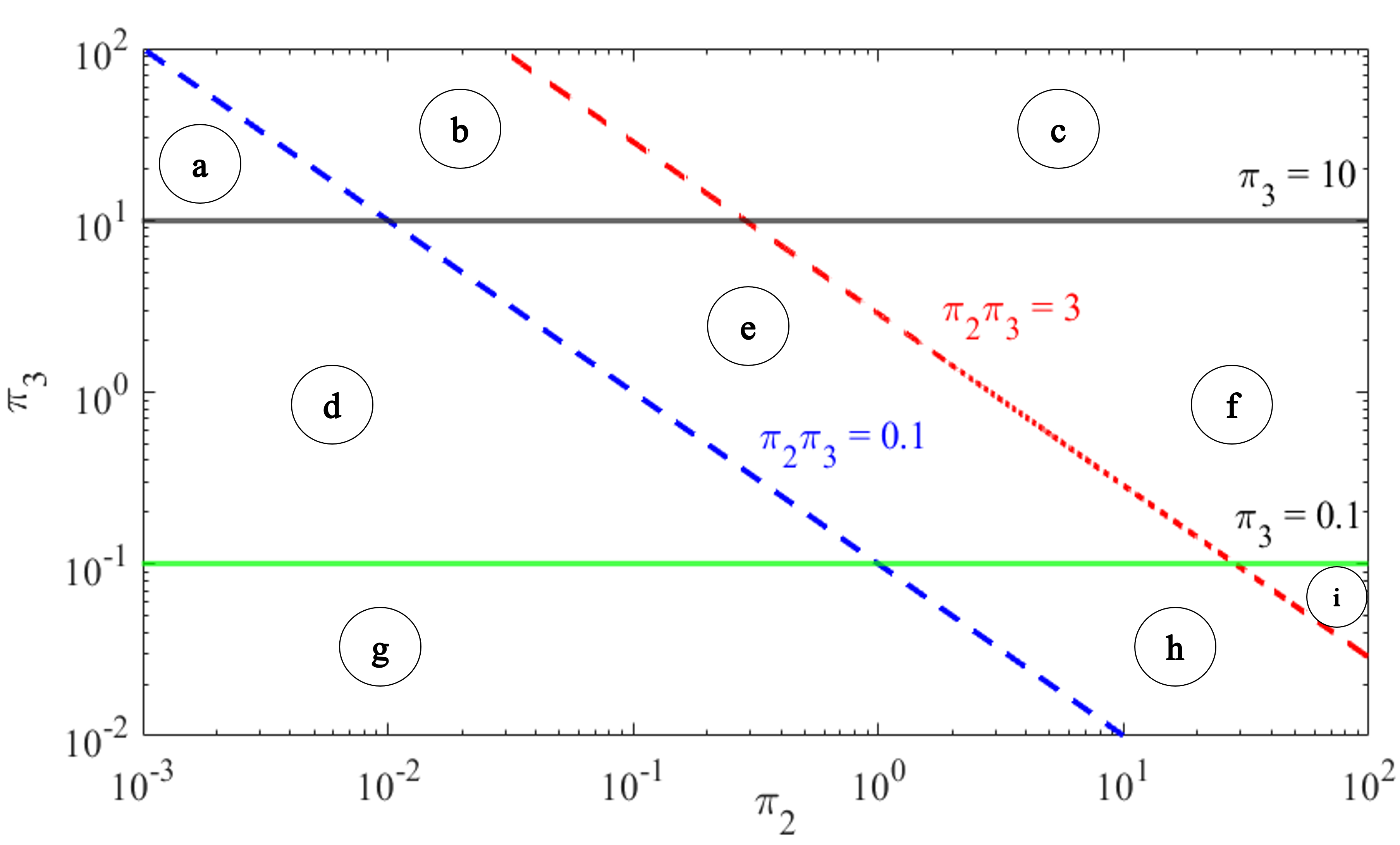}
    \caption{The different regions of operations. The basic constraints $\pi_3 \gg 1$, $\pi_2\pi_3 \ll 1$ and $\pi_3 \ll 1$ are represented as $\pi_3 \ge k$, $\pi_2\pi_3 \le 1/k$ and $\pi_3 \le 1/k$, being $k \gg 1$. Here $k=10$.}
    \label{fig:Feasibility_ext}
\end{figure}

\section{The proposed measurement procedure} \label{section: measurement procedure}
The measurement procedure consists of three main phases subdivided into elementary steps, as outlined in Figure \ref{fig:Op_flowchart}.

The first phase is carried out off-line, once the main parameters such as (i) characteristic of the probe (ii), ranges of interest for the thickness and the electrical conductivity to be estimated, and (iii) the range of interest for the angular frequency, are defined. The first phase is characterized by three tasks: parameters definition, numerical simulations, and experimental calibration. This phase is performed once and for all, given the probe, and the ranges of interest for the electrical conductivity, the thickness and the angular frequencies.

The second phase is in-line and is characterized by two main steps: the experimental execution of the test at a specific angular frequency; the processing of the measured quantity to estimate the thickness and electrical conductivity via the level curves of Figure \ref{fig:Intersection_trend} or \ref{fig:Intersection_trend_norm}. To improve the quality of the thickness and electrical conductivity estimate, Phase 2 can be repeated at different values of the angular frequency.

Finally, in the third phase, the final estimation of the thickness and electrical conductivity is obtained by combining all the level curves from data at different angular frequencies in Figure \ref{fig:Intersection_trend_norm}.

\begin{figure}[htb]
    \centering
    \includegraphics[width=0.70\columnwidth]{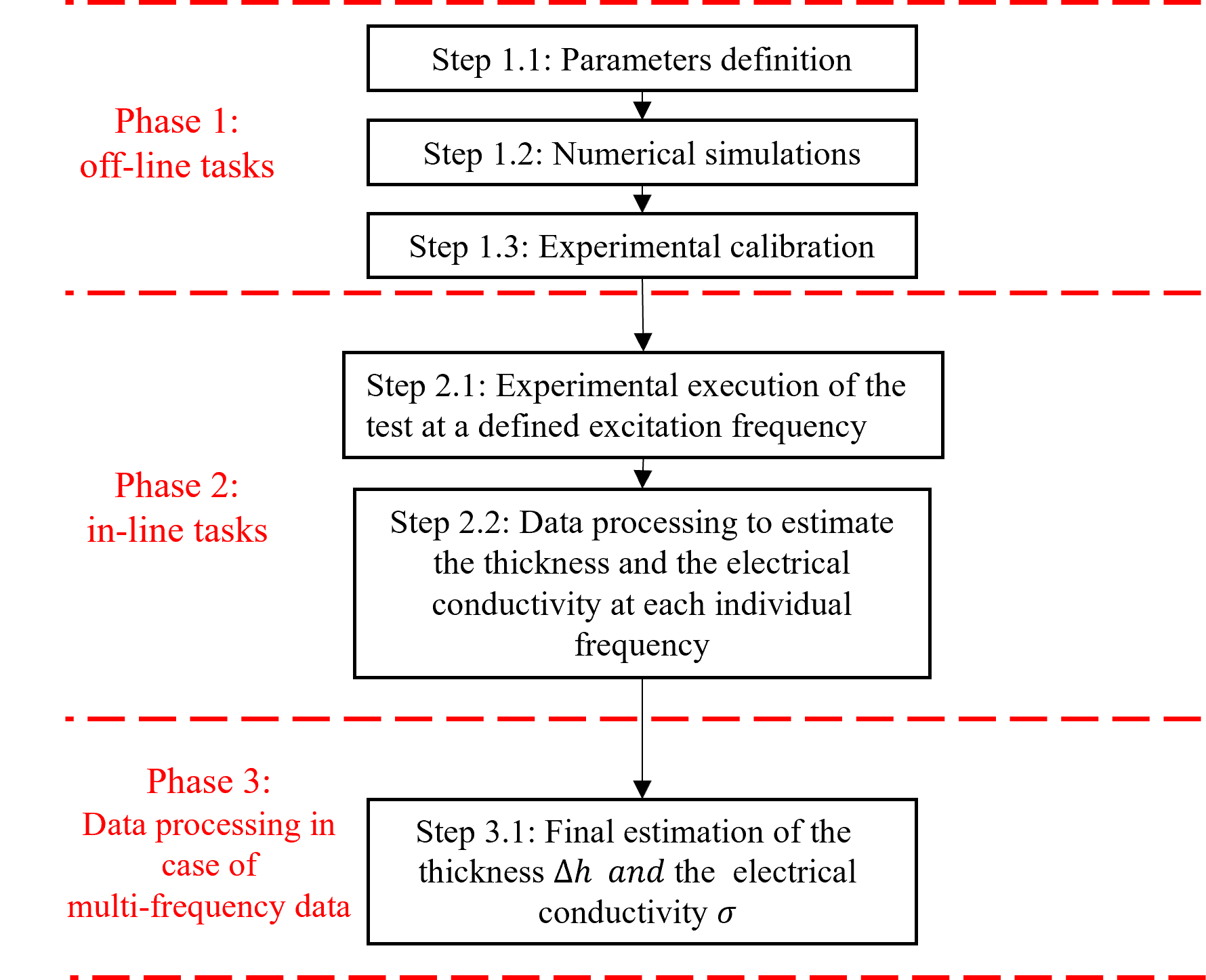}
    \caption{Operation flow-chart behind the $\Delta h - \sigma$ estimation process.}
    \label{fig:Op_flowchart}
\end{figure}

In the following the details for each individual phase are provided.

\textbf{Phase 1: off-line tasks}
\begin{enumerate}
\item[Step \# 1.1] \textit{Parameters definition}
In this task, the characteristics of both the SUT and the ECP are defined. In particular, the following parameters are prescribed: \textit{(i)} geometry, dimensions and physical parameters of the ECP; \textit{(ii)} ranges of interest for both the thickness and the electrical conductivity; \textit{(iii)} range of interest for the driving angular frequency.

The main parameters used in this contribution are summarized in Tables \ref{tab:geometrical_param} and \ref{tab:physical_param}.

\item[Task \# 1.2] \textit{Numerical simulations}
This task entails the numerical computation of the dimensionless impedance, with a view to the calculation of the level curves. This numerical calculation is carried out with reference to the prescribed parameters defined in Task \# 1.1. In detail, for each thickness and frequency value, the corresponding simulated value of $\Delta \dot{Z}_{m}$ is obtained using the semi-analytic models developed by Dodd and Deeds \cite{Dodd_deeds}. Then, the resulting values of the dimensionless groups ($\Bar{\pi}_1$, $\pi_2$, $\pi_3$) are calculated, according to their definitions shown in Table \ref{tab:pi_groups_thick_final}.

During this phase, the key is to avoid under sampling $F(\cdot,\cdot)$ in the region of interest, as guaranteed in the present contribution.

\item[Step \# 1.3] \textit{Experimental calibration}
To process the experimental data via the numerically computed level curves of Step \# 1.2, a suitable calibration is needed. The aim is to get a proper fit between the numerically computed data and the experimentally measured data. As is well known, a mismatch between the numerically computed and the experimental data is due to many factors, such as the specific numerical model, the uncertainty in the knowledge of the geometrical and physical characteristics of the ECP, experimental noise, measurement uncertainty, and so on. In the specific case, by considering a number of different reference plates with known electrical conductivity and thickness, it was found that the ratio $c$ between the numerically computed data and the experimentally measured data depends mainly upon the angular frequency, i.e. $c=c(\omega)$.
\end{enumerate}

\textbf{Phase 2: in-line tasks}
\begin{enumerate}
\item[Task \# 2.1] \textit{Experimental execution of the test at a defined excitation frequency}
This task consists of the experimental measurement of the mutual impedances $\Delta \dot{Z}_{m,k}^{exp}$, at the $k-$th prescribed angular frequency $\omega_k$. The value of $\omega$ is selected from the range defined in Task \# 1.1. This step can be carried out either at a single angular frequency or repeated at other angular frequencies, to improve the accuracy of the estimate of the unknown thickness and electrical conductivity. The SUT is assumed to have an unknown thickness and electrical conductivity within the ranges of Task \# 1.1.

\item[Task \# 2.2] \textit{Data processing to estimate thickness and electrical conductivity at each individual frequency}
First, the calibration factor $c(\omega_k)$ is applied to the mutual impedance $\Delta \dot{Z}_{m,k}^{exp}$ measured at $\omega_k$. Then, the corresponding dimensionless impedance $\Bar{\pi}_{1,k}^{exp}$ is evaluated together with the corresponding level curves for $\Re\{\Bar{\pi}_{1,k}^{exp}\}$, $\Im\{\Bar{\pi}_{1,k}^{exp}\}$, $\Re\{\Bar{\pi}_{1,k}^{exp}\}$ and $|\Bar{\pi}_{1,k}^{exp}|$, and the intersection point $(\pi_{2,k}^{exp},\pi_{3,k}^{exp})$ between the curves. If the position of the intersection point is within one of the regions of interest (d), (e) or (h) of Figure \ref{fig:Feasibility_ext}, then the $k-$th estimate of the electrical conductivity and thickness is given by $\Delta h_k = D \pi_{3,k}^{exp}$ and $\sigma_k =\frac{2 \nu_0}{\omega_k} \left( \frac{\pi_{2,k}^{exp}}{D} \right)^2$.

Task 2 is repeated at a different angular frequencies if either $(\pi_{2,k}^{exp},\pi_{3,k}^{exp})$ is not within one of the region of interest (d), (e) or (h) of Figure \ref{fig:Feasibility_ext}, or to improve the accuracy of the estimate of $\sigma$ and $\Delta h$, by leveraging measurements from different frequencies.
\end{enumerate}

\textbf{Phase 3: Data processing for thickness and electrical conductivity estimation in a multi-frequency approach}
\begin{enumerate}
\item[] This last task is carried out only for measurements at multiple angular frequencies to improve the accuracy of the estimate of $\sigma$ and $\Delta$. The final estimate can be achieved by combining level curves from different angular frequencies or, more simply, by averaging the $\sigma_k$s and $\Delta h_k$s estimated at each individual angular frequency $\omega_k$. Hereafter, for the sake of simplicity, the results arising from the latter choice are shown.
\end{enumerate}

\section{Experimental characterization of the proposed measurement method} \label{section:experimental_charact}
\subsection{Experimental set--up} \label{subsection:set_up}
The experimental set-up consists of an ECP, a waveform generator, a current probe, two signal amplifiers, a data acquisition board and a Personal Computer (PC). A schematic block diagram of the experimental set-up is shown in Figure \ref{fig:Set_up}.

\begin{figure}[htb]
    \centering
    \includegraphics[width=0.60\columnwidth]{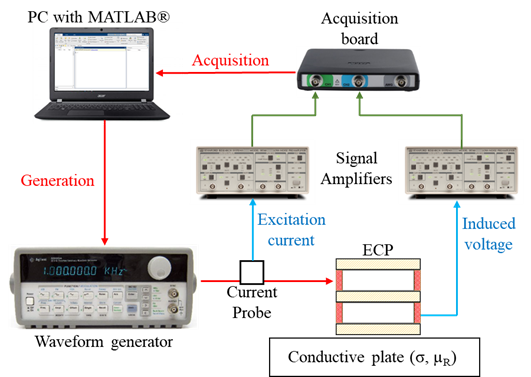}
    \caption{Schematic block diagram of the experimental set-up.}
    \label{fig:Set_up}
\end{figure}

The ECP consists of two coaxial coils, a driving coil and a receiver coil. The geometry of the coils and their dimensions are provided in Figure \ref{fig:Probe} and Table \ref{tab:geometrical_param}, respectively. An Agilent 33120A waveform generator provides the source current to the ECP driving coil, which is sensed by means of a Tektronix TCP202A current probe. Both the output of the current probe, proportional to the driving current, and the output voltage at the receiver coil, are conditioned by means of two SR560 Stanford Research System low-noise signal amplifiers. A TIE-PIE Engineering Handyscope $HS5-540XMS-W5^{TM}$ data acquisition board is adopted to digitize the output signals provided by the two conditioning units (the signal amplifiers).

The experimental tests were carried out at several angular frequencies, to perform a metrological characterization of the method. Specifically, the frequencies from $300$ Hz to $3$ kHz, with a frequency step equal to $50$ Hz, were considered. The RMS value of the applied current at each angular frequency was $115$ mA. The conditioning unit consists of a bandpass filter from $30$ Hz to $10$ kHz, a gain of $200$ for the output of the current probe and a gain of $100$ for the output measured at the receiver coil. To optimize the data processing in the time domain, signal digitization is performed by adopting a sampling frequency $1000$ times the considered signal frequency and acquiring $4$ periods of each signal. A script developed in the MATLAB$^{TM}$, running on a dual-core PC, manages the automation of the measurement station and performs the signal processing. The final output of the measurement station is the mutual impedance $\Delta \dot{Z}_{m,k}^{exp}$.

\subsection{Results and discussion} \label{subsection:experimental_results}
The experimental set-up described in Section \ref{subsection:set_up} is used to make both the off-line tasks (Step \# 1.3) and the in-line tasks (Step \# 2.1). As regards Step \# 2.1, the tests were carried out on six plates with known thicknesses and electrical conductivities. The main characteristics of the plates are listed in Table \ref{tab:plates_param}.

\begin{table}[htb]
\captionsetup{justification=centering}
\caption{Main characteristics of analyzed plates.}
\begin{adjustbox}{width=\columnwidth,center}
\begin{tabular}{cccccl}
\hline \hline
Name code    & Metal alloy                       & Electrical conductivity ($\widetilde{\sigma}$) {[}MS/m{]} & Thickness ($\widetilde{\Delta h}$) {[}mm{]} & Plate dimensions {[}mm x mm{]} &  \\ \hline
$\#$a       & Aluminium (2024-T3)                & 17.66                                      & 2.03                     & {200 x 200}                    &  \\
$\#$b       & Copper                             & 58.50                                      & 0.98                      & {200 x 200}                    &  \\
$\#$c       & Aluminium (6061-T6)                & 28.23                                      & 1.97                      & {200 x 200}                    &  \\
$\#$d       & Aluminium (AW-1050A)               & 35.27                                      & 1.03                      & {250 x 250}                    &  \\
$\#$e       & Aluminium (AW-1050A)               & 35.44                                      & 2.93                      & {250 x 250}                    &  \\ 
$\#$f       & Aluminium (AW-1050A)               & 35.91                                      & 3.98                      & {250 x 250}                    &  \\  \hline \hline
\label{tab:plates_param}
\end{tabular}
\end{adjustbox}
\end{table}

For each specific plate and driving angular frequency, 20 repeated measurements were carried out to investigate repeatability in the estimation of both electrical conductivity and the thickness.

\begin{figure}[htp]
    \centering
    \subfloat[][]
    {\includegraphics[width=.45\textwidth]{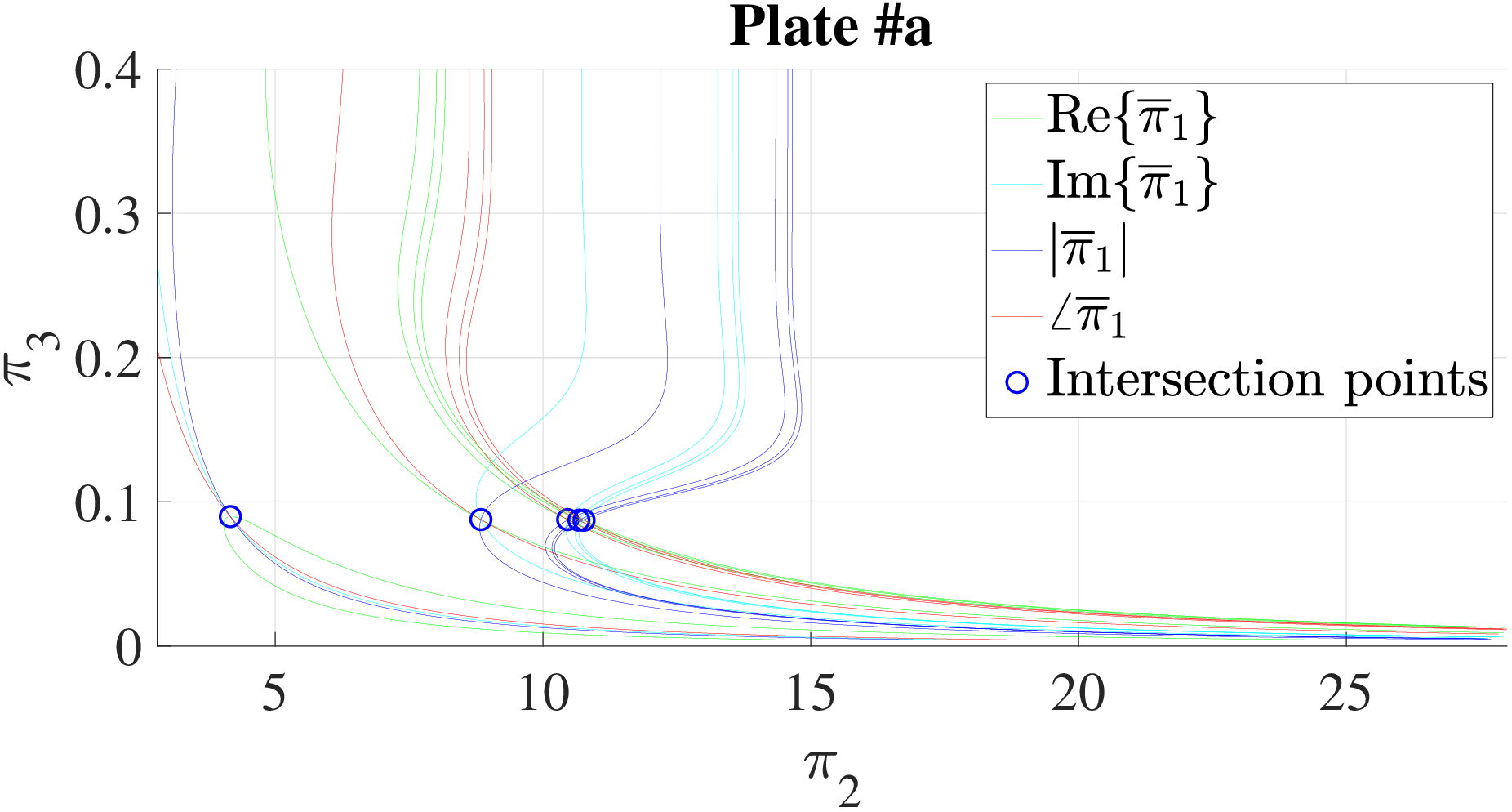}} \quad
    \subfloat[][]
    {\includegraphics[width=.45\textwidth]{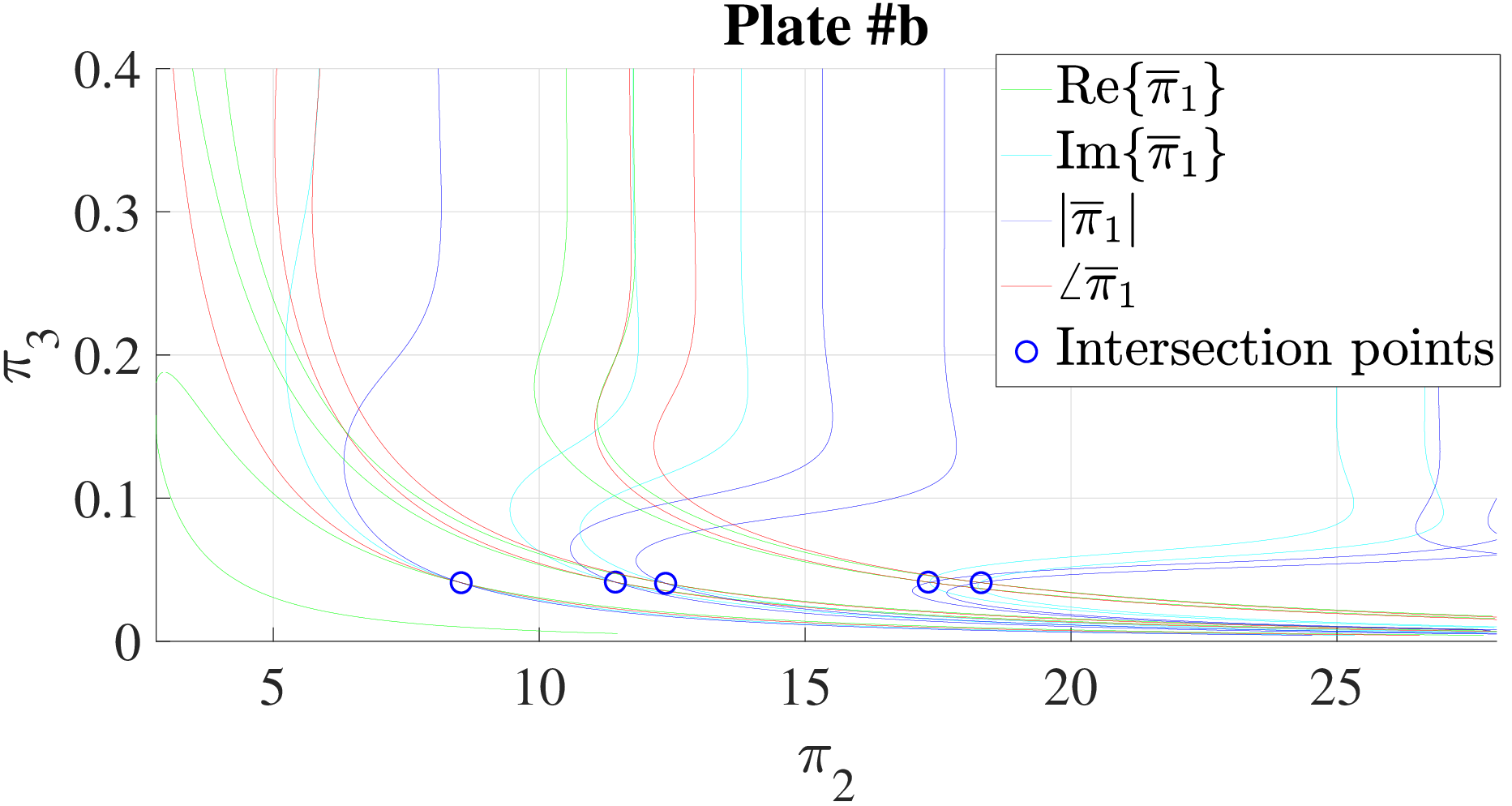}} \\
    \subfloat[][]
    {\includegraphics[width=.45\textwidth]{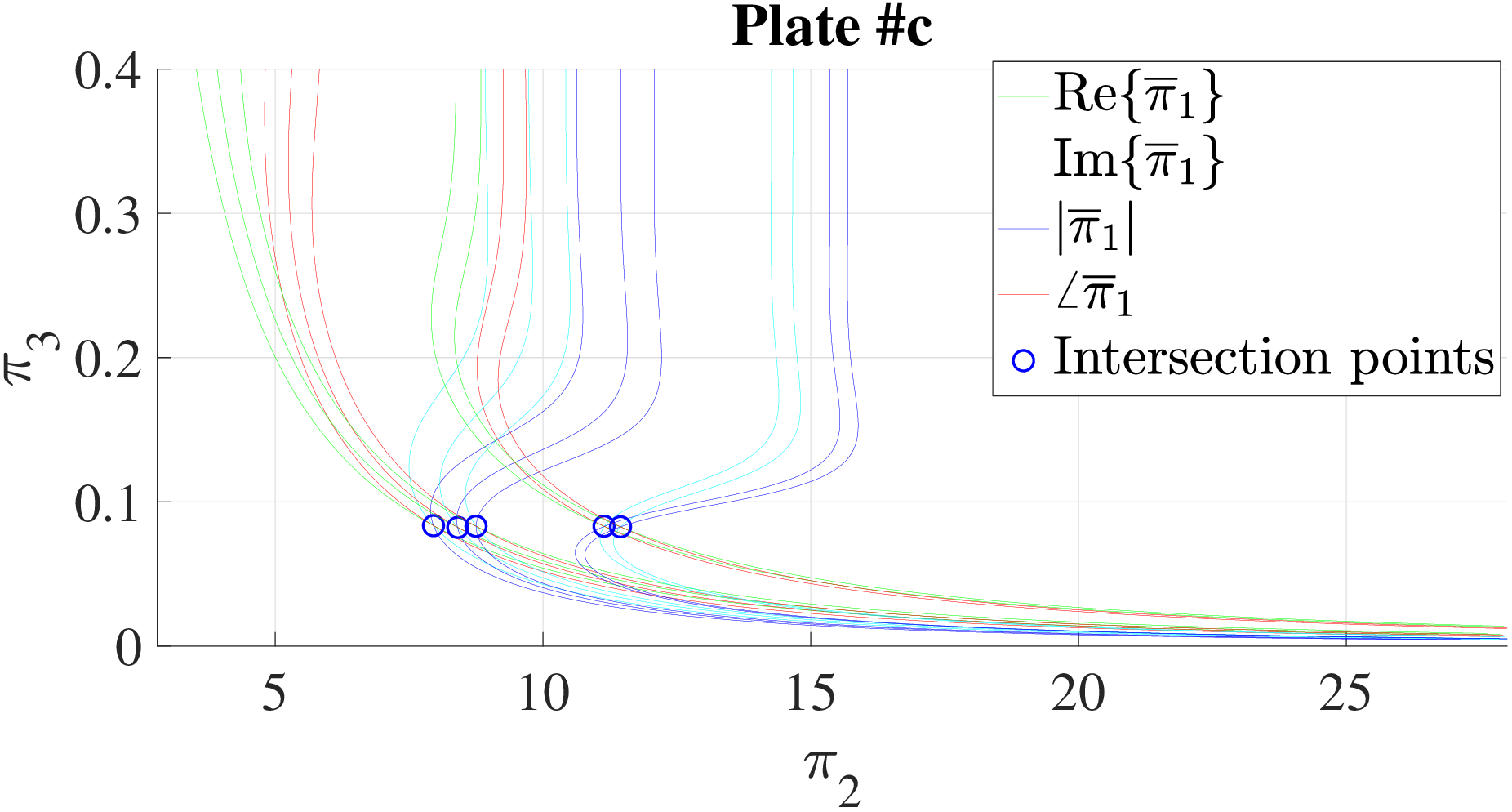}} \quad
    \subfloat[][]
    {\includegraphics[width=.45\textwidth]{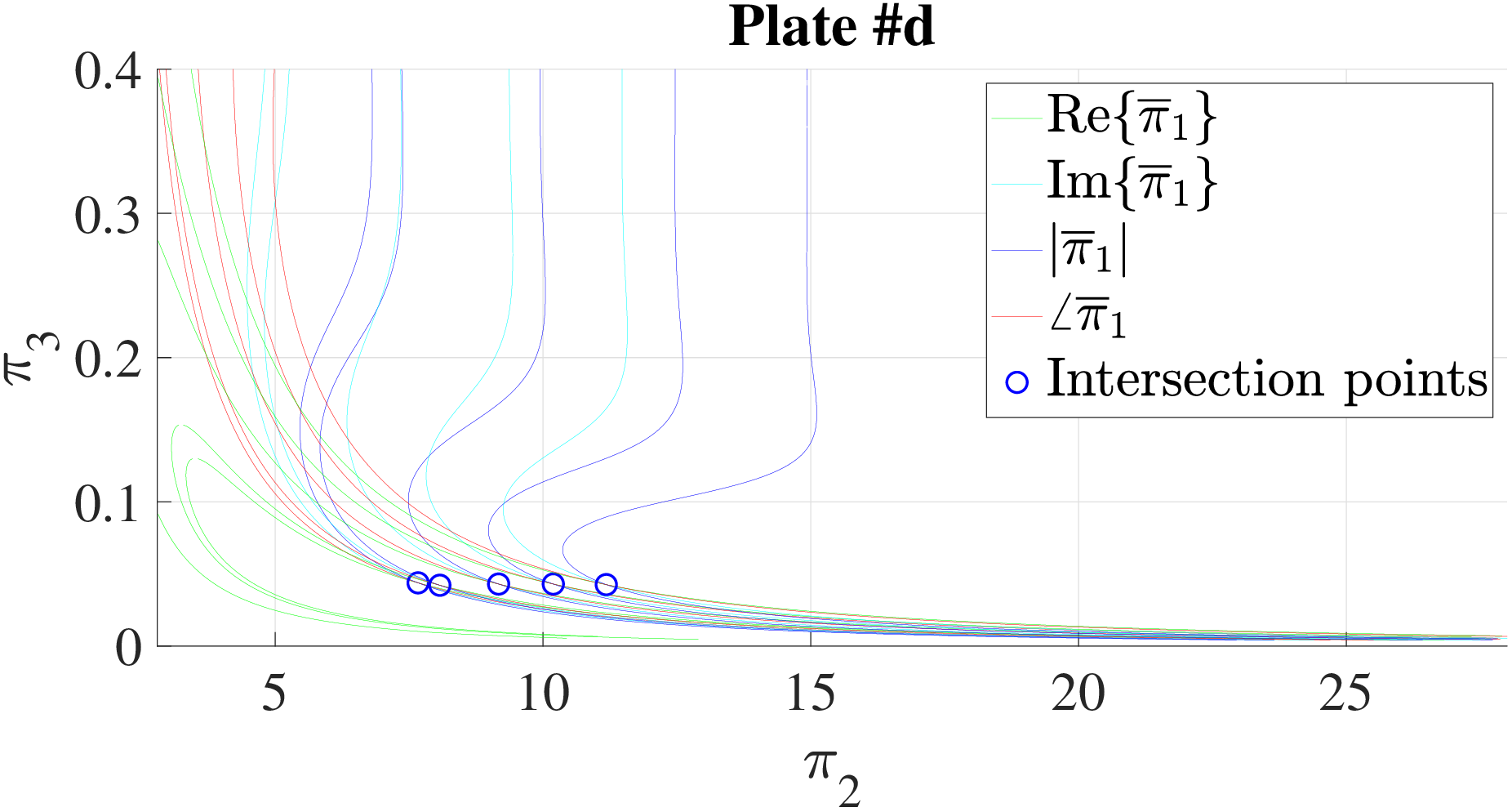}} \\
    \subfloat[][]
    {\includegraphics[width=.45\textwidth]{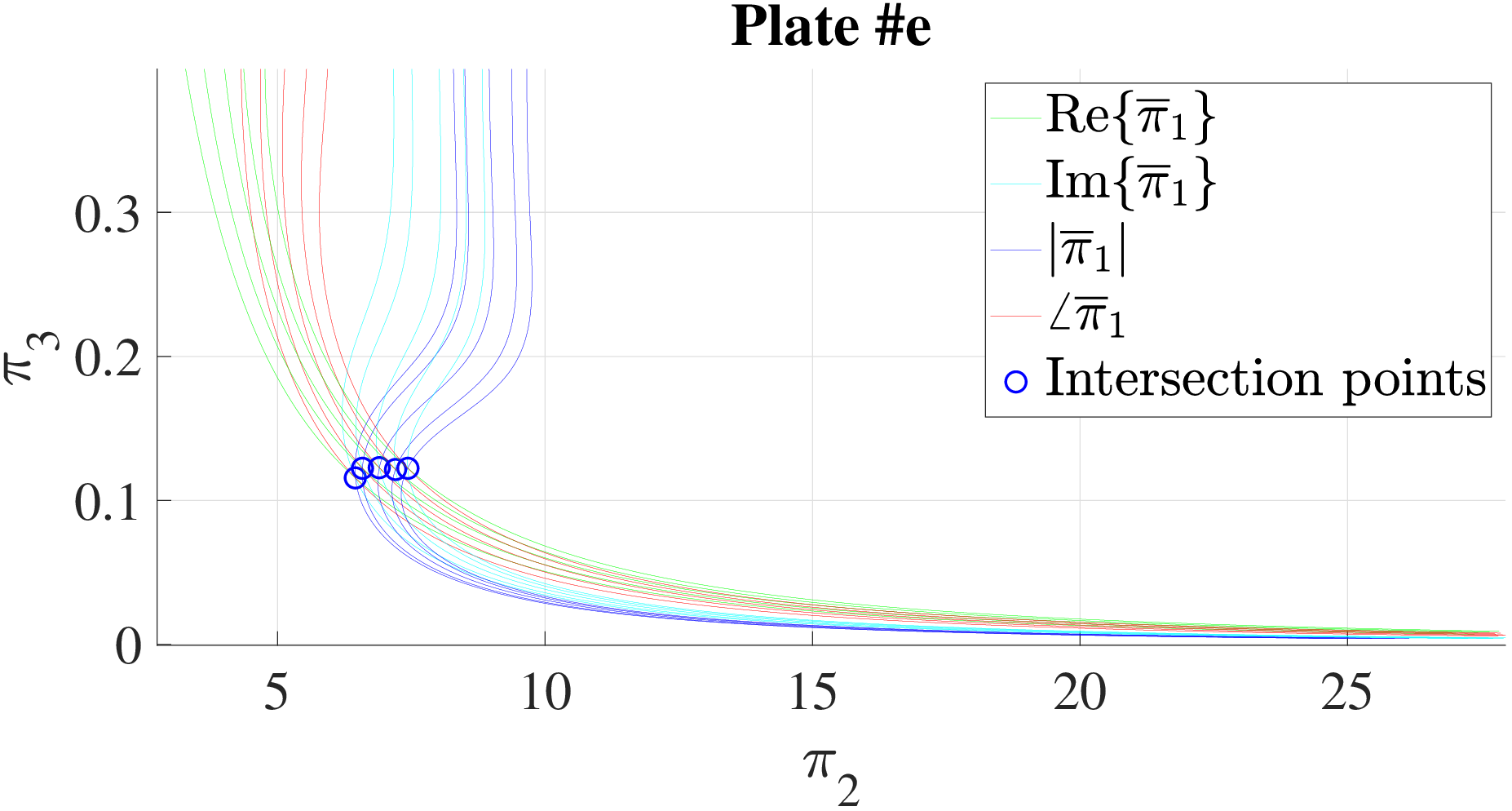}} \quad
    \subfloat[][]
    {\includegraphics[width=.45\textwidth]{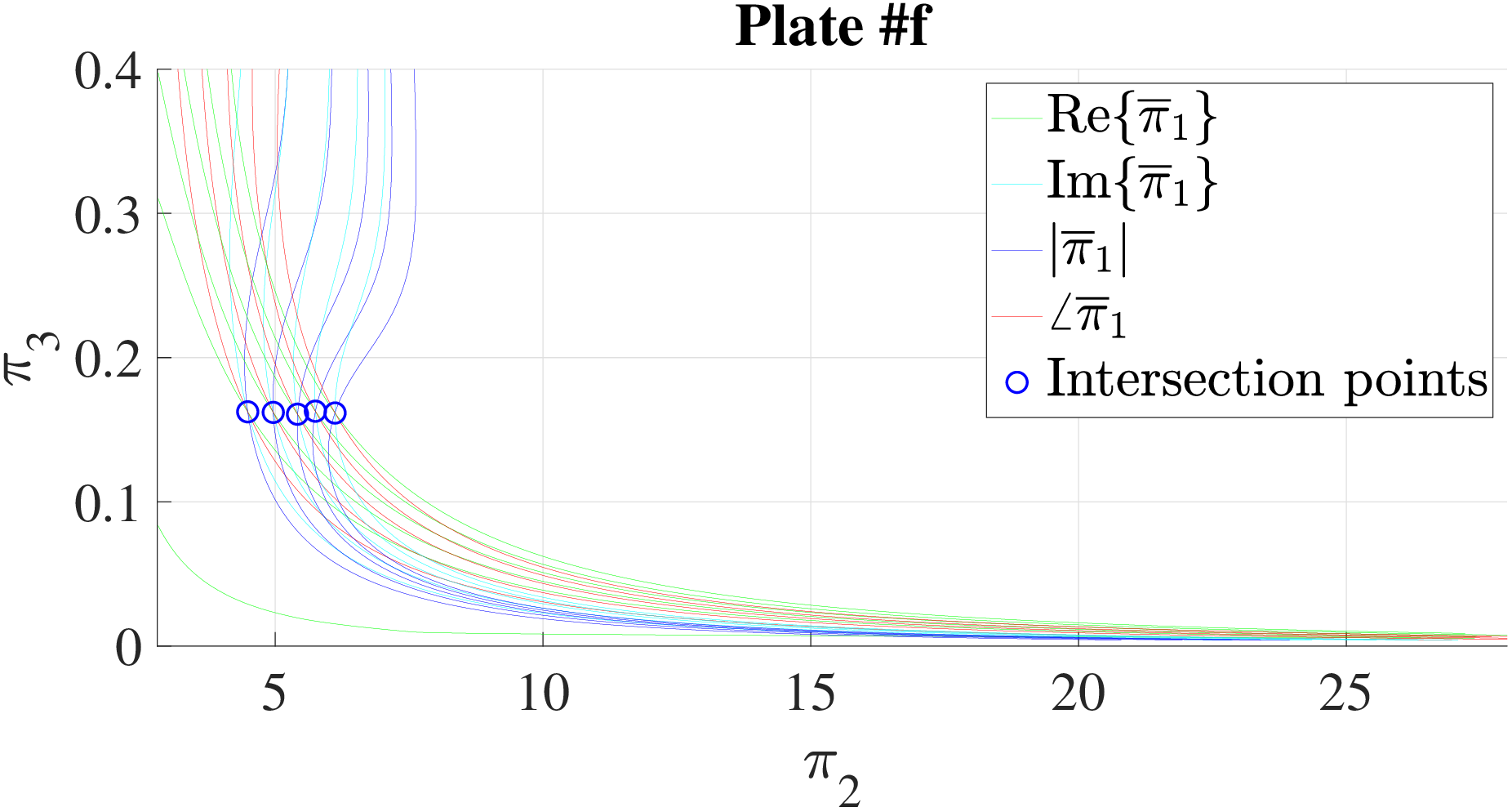}} 
    \caption{Level curves corresponding to the experimental data for all six plates in the $(\pi_2,\pi_3)$ plane, at the five angular frequencies corresponding to the lowest estimation errors.}
    \label{fig:Intersect_pi1_pi2_Exp}
\end{figure}

\begin{figure}[htp]
    \centering
    \subfloat[][]
    {\includegraphics[width=.45\textwidth]{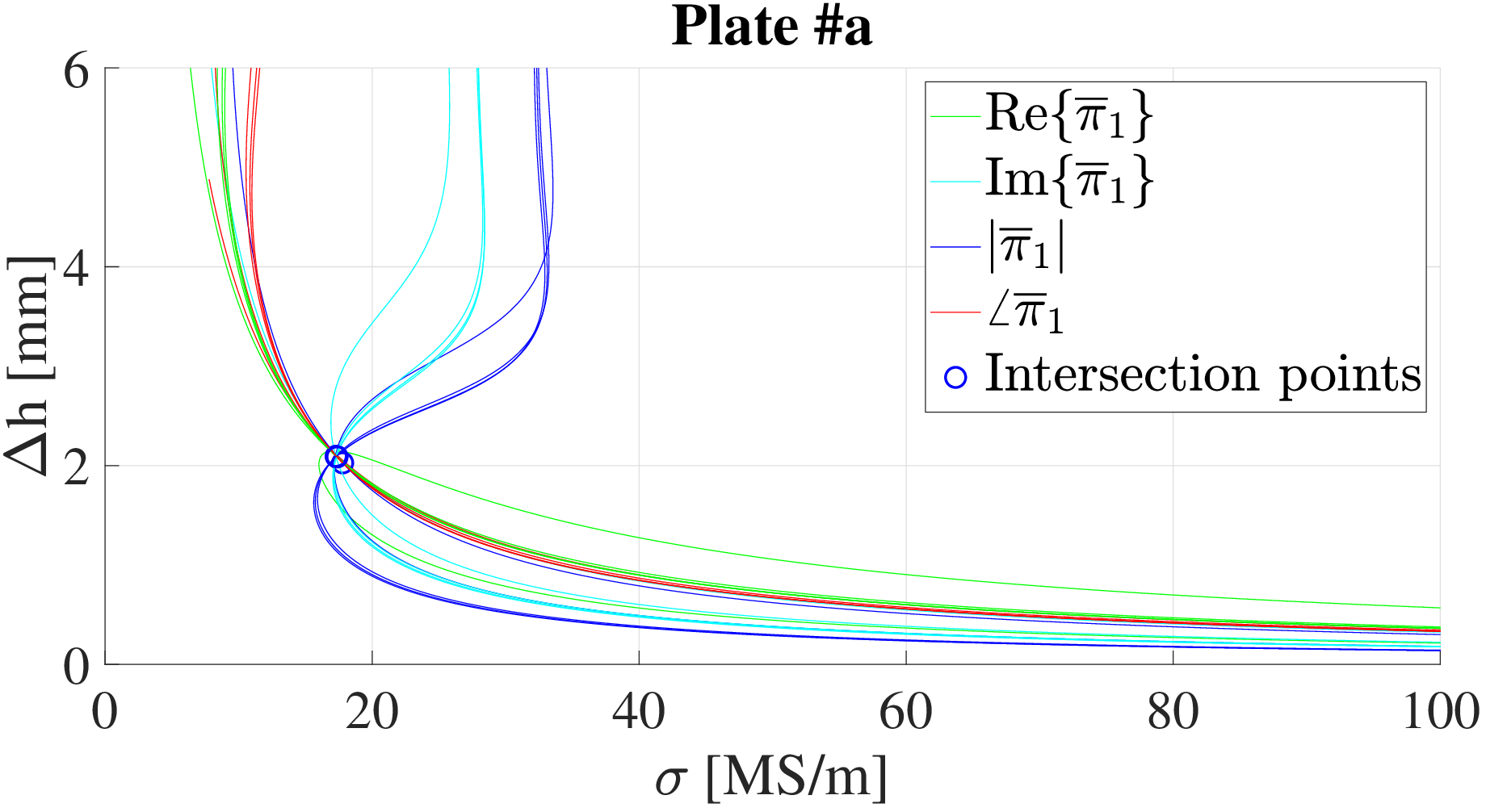}} \quad
    \subfloat[][]
    {\includegraphics[width=.45\textwidth]{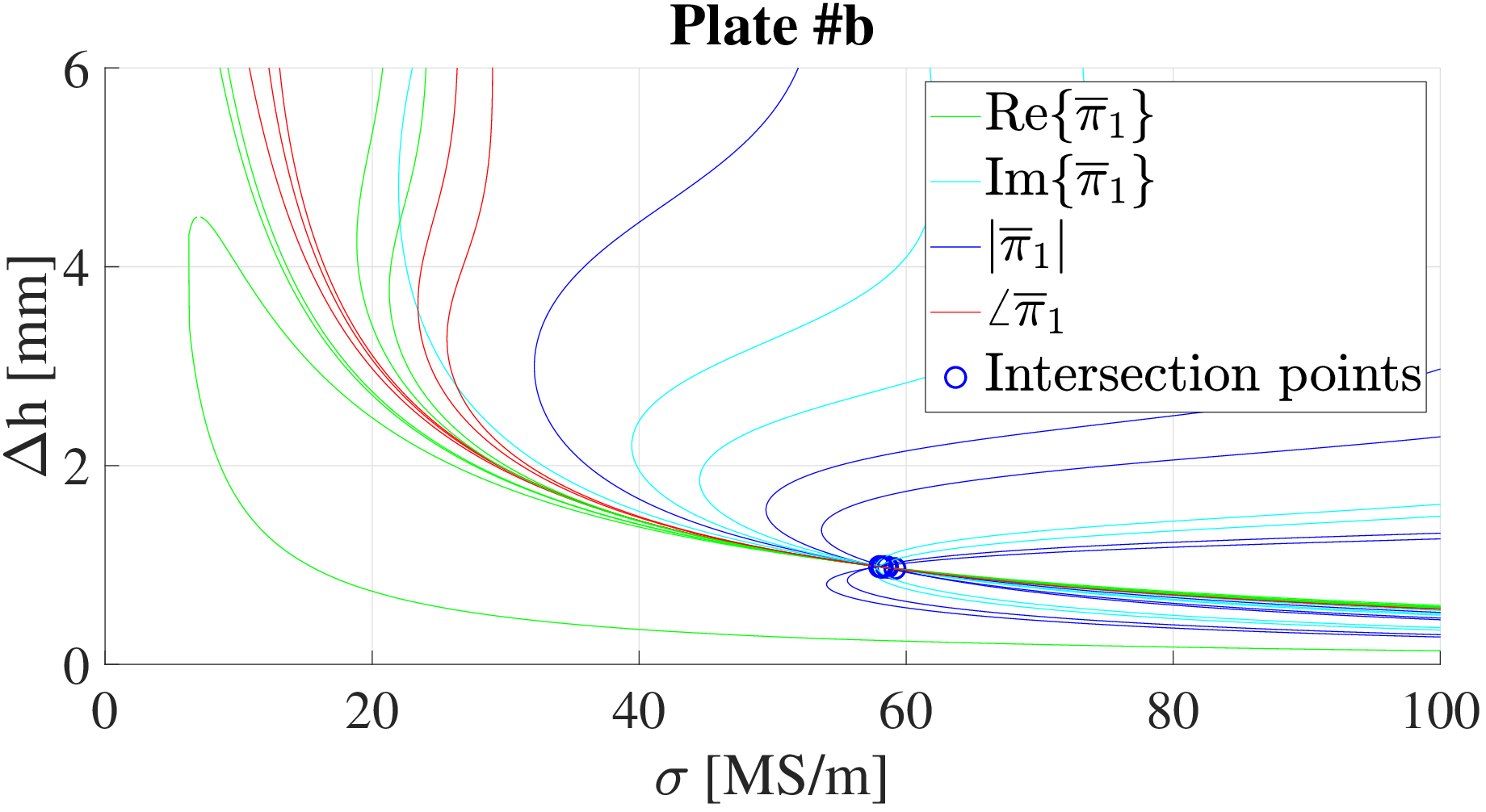}} \\
    \subfloat[][]
    {\includegraphics[width=.45\textwidth]{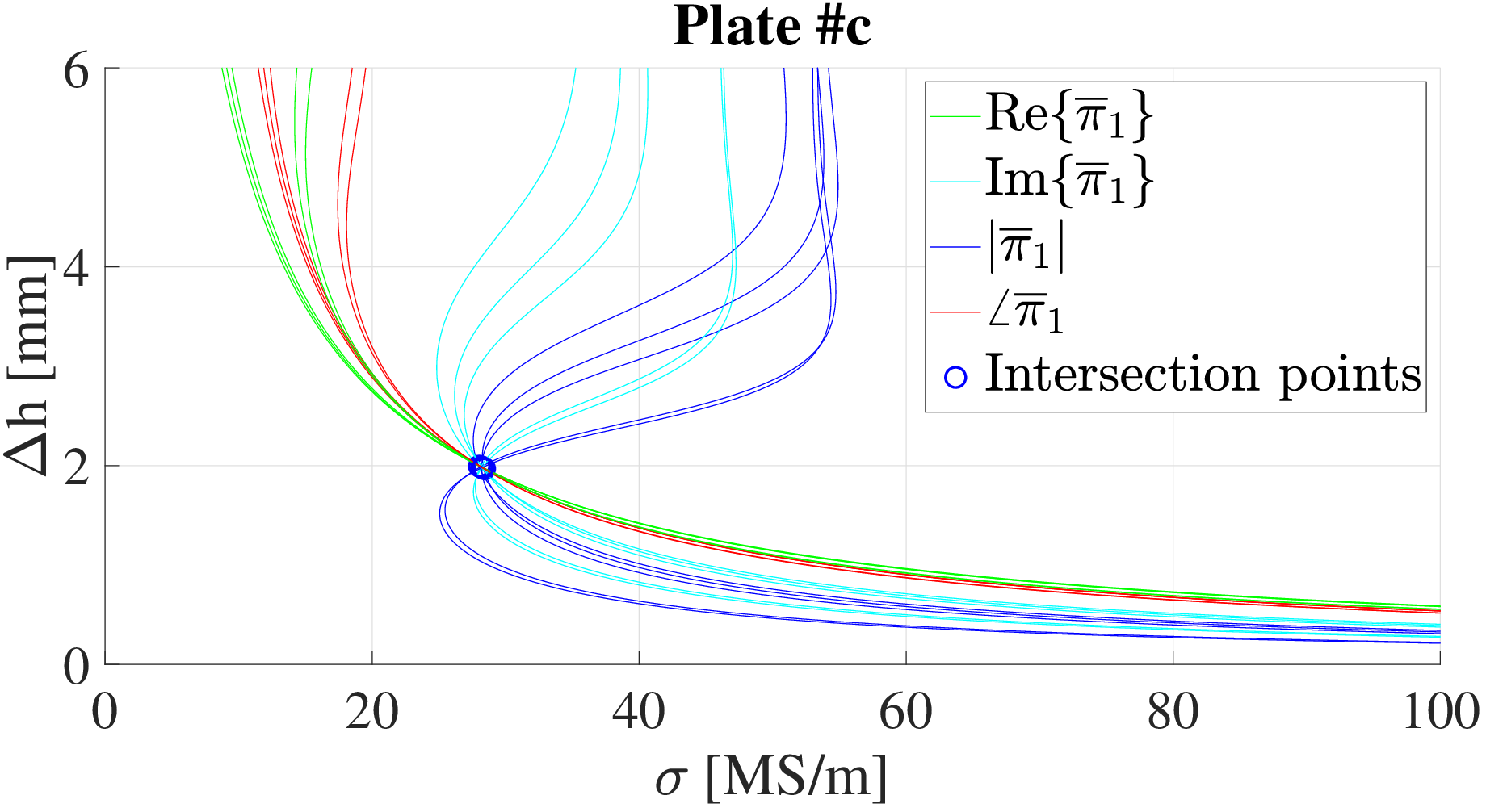}} \quad
    \subfloat[][]
    {\includegraphics[width=.45\textwidth]{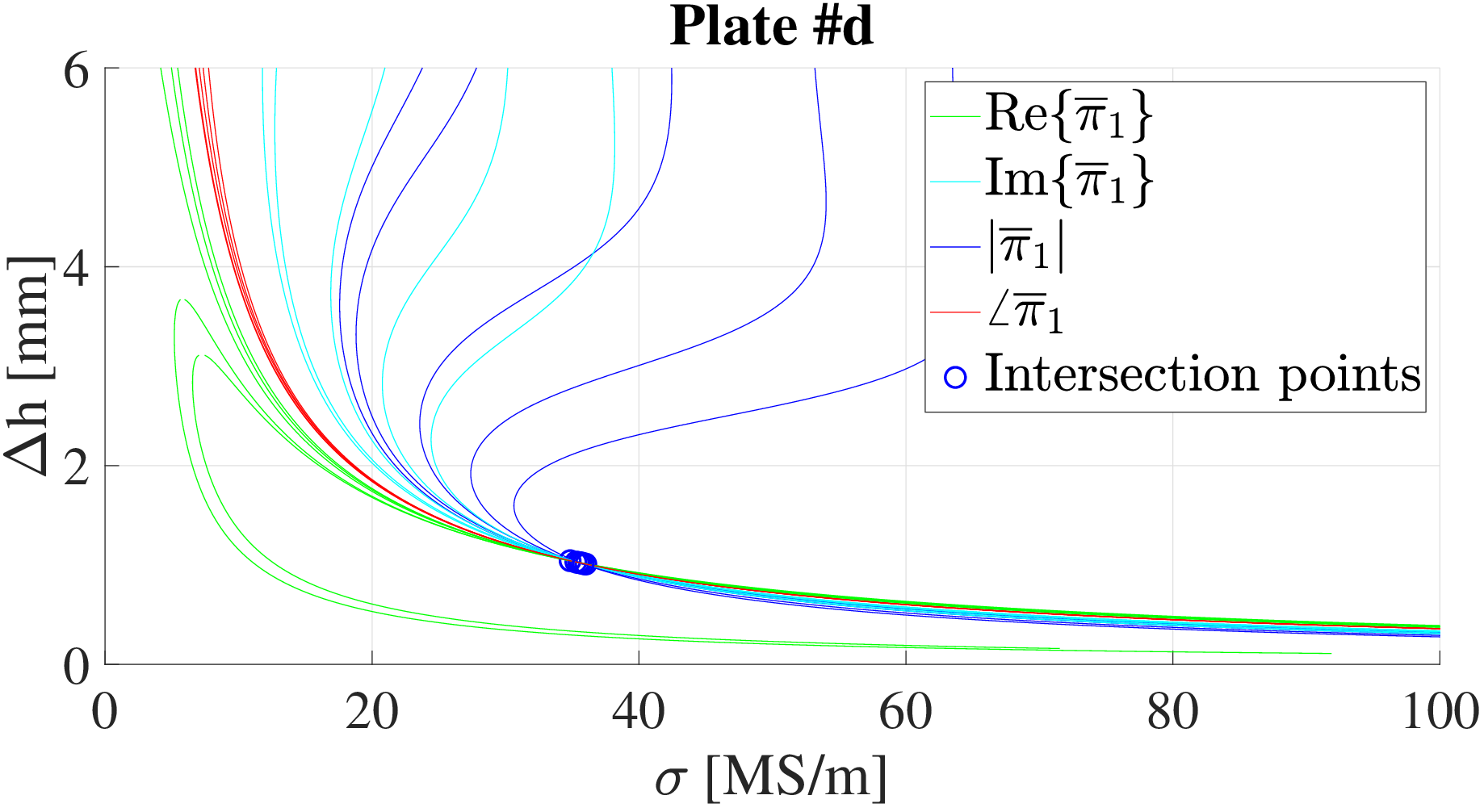}} \\
    \subfloat[][]
    {\includegraphics[width=.45\textwidth]{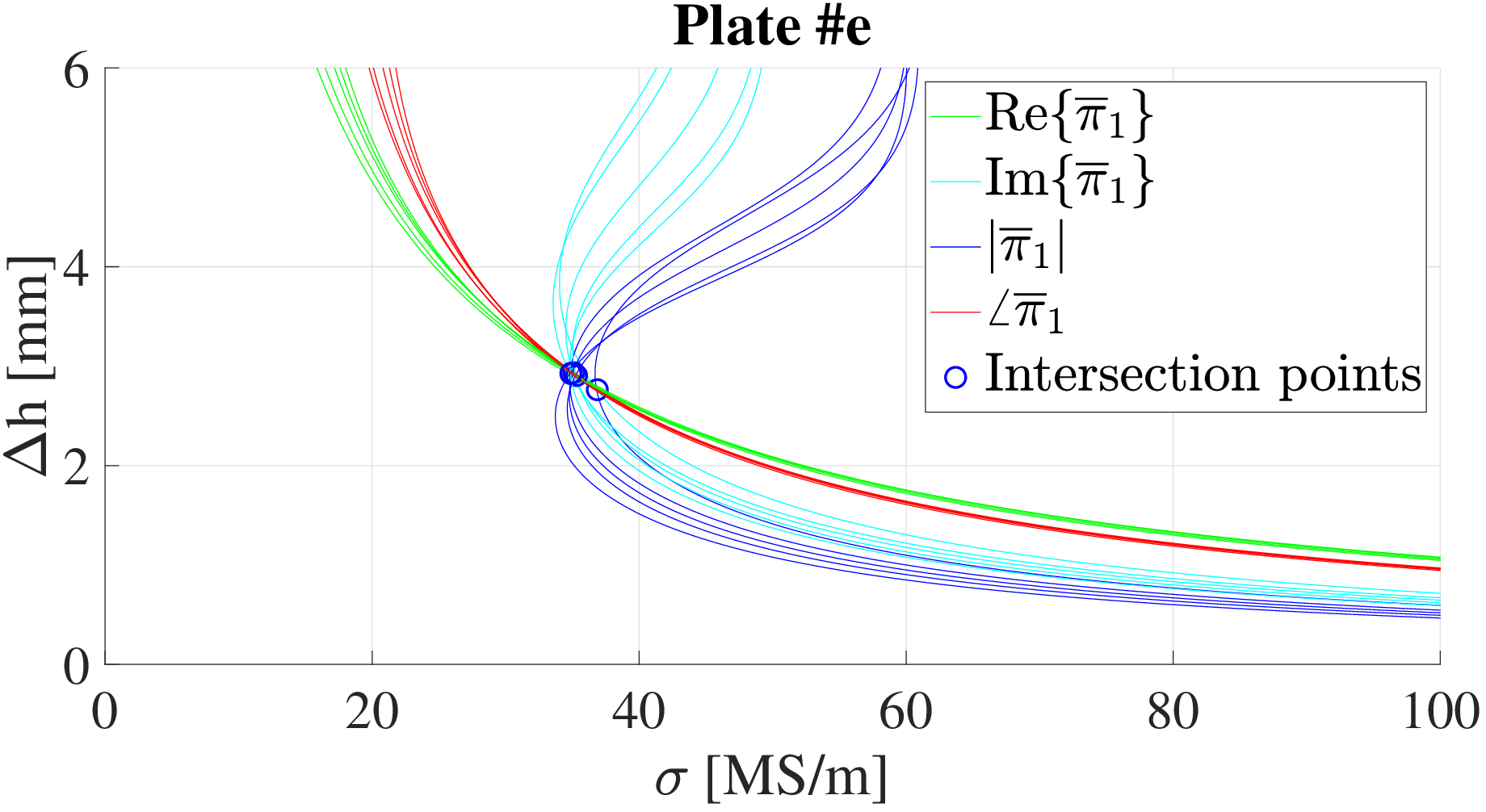}} \quad
    \subfloat[][]
    {\includegraphics[width=.45\textwidth]{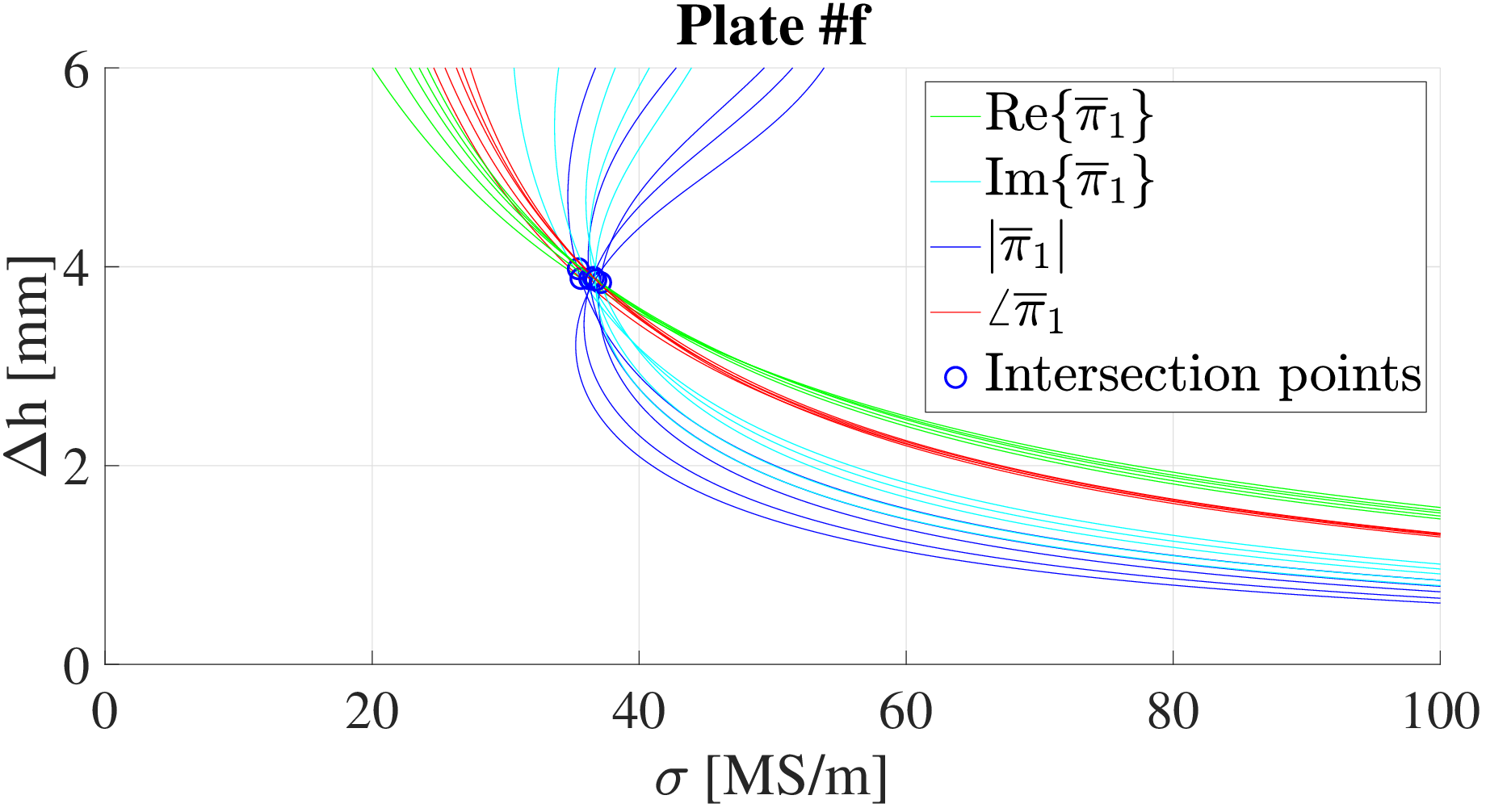}} 
    \caption{Level curves corresponding to the experimental data for all six plates in the $(\sigma,\Delta h)$ plane, at the five angular frequencies corresponding to the lowest estimation errors.}
    \label{fig:Intersect_st_Exp}
\end{figure}

Figures \ref{fig:Intersect_pi1_pi2_Exp} and \ref{fig:Intersect_st_Exp} provide some examples of level curves corresponding to the experimental data for all six plates, in both the $(\pi_2,\pi_3)$ and the $(\sigma,\Delta h)$ planes. To each frequency corresponds one single intersection of the level curves for the real part, imaginary part, magnitude and phase of the dimensionless impedance. The position of the intersections of the curves in the $(\pi_2,\pi_3)$ plane makes it possible to establish whether or not data has to be retained. Indeed, when estimating $\sigma$ and $\Delta h$, if the intersection does not fall in region (d), (e) or (h) of Figure \ref{fig:Feasibility_ext}, the data has to be discarded. This criterion is very powerful because it can be applied without \emph{any} knowledge of the unknown parameters $\sigma$ and $\Delta h$. The position of the intersections in the $(\sigma,\Delta h)$ makes it possible to evaluate the accuracy of the estimate. Ideally, all intersections should be located in a single point. When this is not the case, the spreading of the intersections gives the accuracy of the estimate of the unknown quantities. Points that are \lq\lq far\rq\rq \ from the main cluster, can be retained as outliers and, therefore, safely discarded.

Hereafter, we define $\sigma_k^i$ as the $i-th$ estimate of the electrical conductivity at the $k-th$ angular frequency, where $i=1,\ldots,20$ (20 repeated measurements). Similarly, we define $\Delta h_k^i$ as the $i-th$ estimate of the thickness at the $k-th$ angular frequency.

To quantitatively analyze the performances levels of the proposed method, the following figures of merit have been considered:
\begin{itemize}
    \item The average electrical conductivity $\overline{\sigma}_k$ and the average thickness $\overline{\Delta h}_k$, obtained from the 20 repeated measurements at the $k-$th angular frequency.
    \item The absolute relative errors $\epsilon_{\sigma,k}$ and $\epsilon_{\Delta h,k}$ for the mean value of electrical conductivity and thickness, respectively:
    \begin{equation}
        \epsilon_{\sigma,k} = \frac{|\overline{\sigma}_k-\widetilde{\sigma}|}{\widetilde{\sigma}} \cdot 100, \ \epsilon_{\Delta h,k} = \frac{|\overline{\Delta h}_k-\widetilde{\Delta h}|}{\widetilde{\Delta h}} \cdot 100.
    \label{eq:relative_error_sigma}
    \end{equation}
    In \eqref{eq:relative_error_sigma} $\widetilde{\sigma}$ and $\widetilde{\Delta h}$ are the known values of electrical conductivity and thickness, respectively.
    \item The standard deviations $std_{\epsilon_{\sigma,k}}$ and $std_{\epsilon_{\Delta h,k}}$  for the sets $\{\epsilon_{\sigma,k}^1,\ldots,\epsilon_{\sigma,k}^{20}\}$ and $\{\epsilon_{\Delta h,k}^1,\ldots,\epsilon_{\Delta h,k}^{20}\}$, where
    \begin{equation}
        \epsilon_{\sigma,k}^i = \frac{|{\sigma}_k^i-\widetilde{\sigma}|}{\widetilde{\sigma}} \cdot 100, \ \epsilon_{\Delta h,k}^i = \frac{|{\Delta h}_k^i-\widetilde{\Delta h}|}{\widetilde{\Delta h}} \cdot 100.
    \label{eq:relative_error_i_k}
    \end{equation}
    \item The final estimate of electrical conductivity $\overline{\sigma}$ and of thickness $\overline{\Delta h}$, obtained by averaging the $\overline{\sigma}_k$s and the $\overline{\Delta h}_k$s, respectively.
    \item The relative error for the final estimate $\overline{\sigma}$ and $\overline{\Delta h}$, with respect to $\widetilde{\sigma}$ and $\widetilde{\Delta h}$:
    \begin{equation}
    \epsilon_{\sigma} = \frac{|\overline{\sigma}-\widetilde{\sigma}|}{\widetilde{\sigma}} \cdot 100, \         \epsilon_{\Delta h} = \frac{|\overline{\Delta h}-\widetilde{\Delta h}|}{\widetilde{\Delta h}} \cdot 100
    \label{eq:relative_error_H_media}
    \end{equation}    
    \item The standard deviations $std_{\sigma}$ and $std_{\Delta h}$ for the $\overline{\sigma}_k$s and $\overline{\Delta h}_k$s.
    \item The standard deviations $std_{\epsilon_{\sigma}}$ and $std_{\epsilon_{\Delta h}}$ for the $\epsilon_{\sigma,k}$s and the $\epsilon_{\Delta h,k}$s.
 \end{itemize}

Figure \ref{fig:freq_deltaH} shows the behaviour $\epsilon_{\Delta h,k}$ and $std_{\epsilon_{\Delta h,k}}$, related to the thickness estimate. The figures of merit for estimating electrical conductivity are shown in Figure \ref{fig:freq_sigma}. The frequency range for different plates varies, to guarantee operations within regions (d), (e) and (h), as discussed in Section \ref{subsection:operating_area} and Figure \ref{fig:Feasibility}.

\begin{figure}[htb]
    \centering
    \includegraphics[width=0.75\columnwidth]{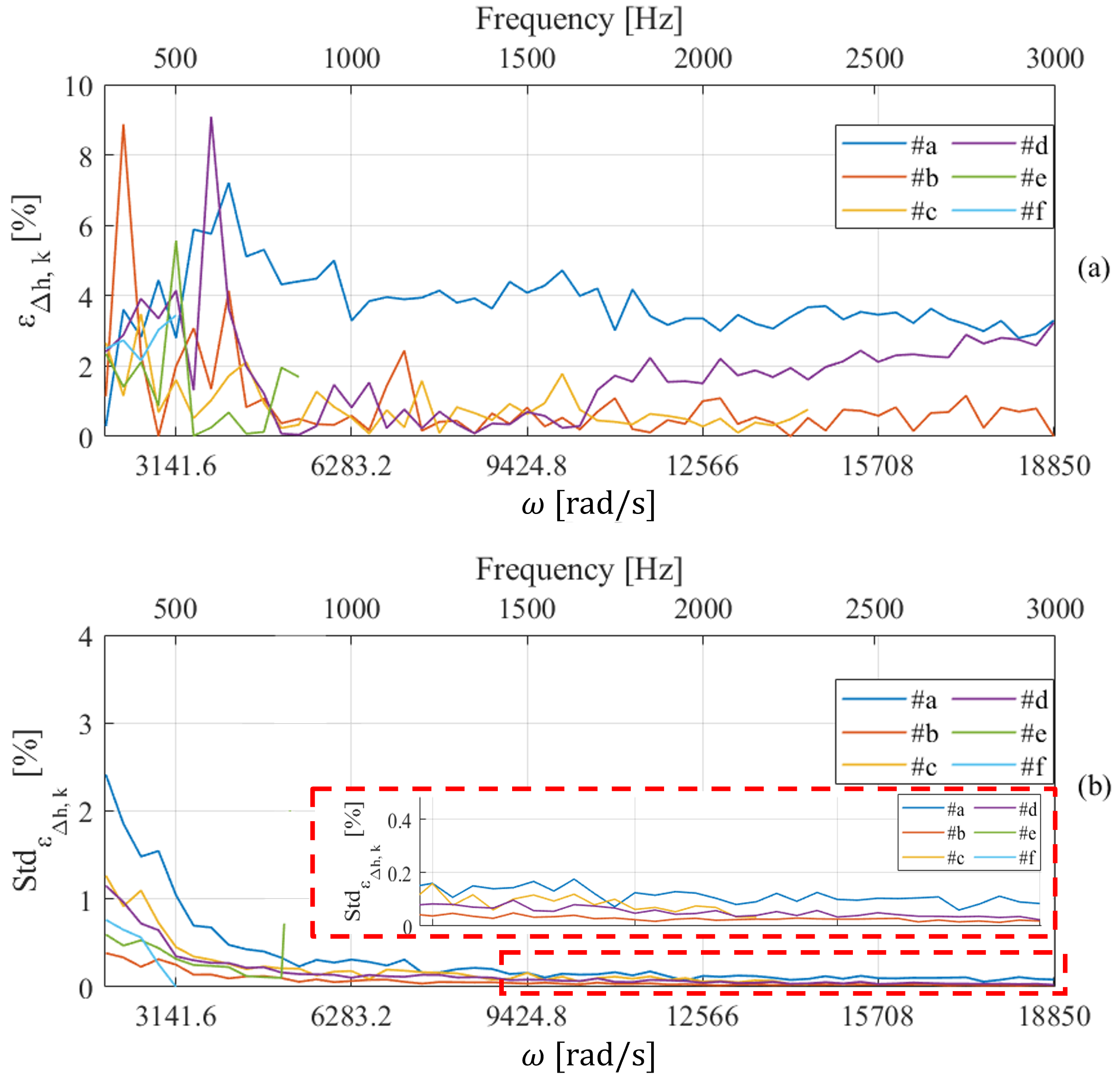}
    \caption{(a) Behaviour of the mean relative error ($\epsilon_{\Delta h, k}$) and (b) the corresponding relative standard deviation ($std_{\epsilon_{\Delta h, k}}$) for the estimated thicknesses at different frequencies and for all the considered metallic plates.}
    \label{fig:freq_deltaH}
\end{figure}

\begin{figure}[htb]
    \centering
    \includegraphics[width=0.75\columnwidth]{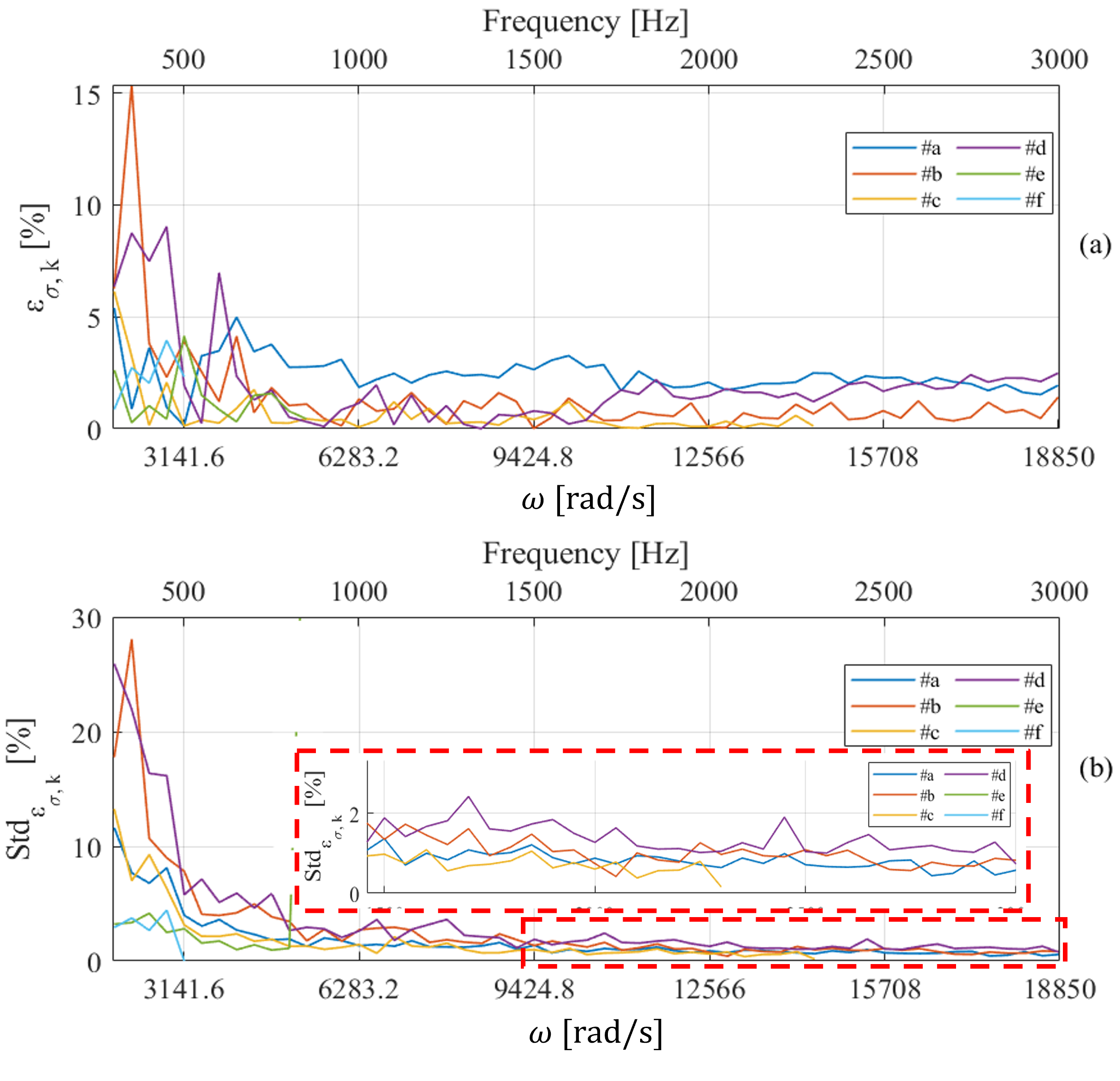}
    \caption{(a) Behaviour of the mean relative error ($\epsilon_{\sigma, k}$) and (b) the corresponding relative standard deviation ($std_{\epsilon_{\sigma, k}}$) for the estimated electrical conductivities at different frequencies and for all the considered metallic plates.}
    \label{fig:freq_sigma}
\end{figure}

As expected, for all the analysed plates, both the error and the standard deviations are generally more significant at low excitation frequencies for both $\Delta h$ and $\sigma$, due to the weakness of the intensity of the eddy currents at those angular frequencies. As the driving frequency is increased, the relative error assumes suitable values lower than 5 \% for thickness and 4 \% for electrical conductivity. Similar behaviors can be observed for the standard deviations (see Figure \ref{fig:freq_deltaH} (b) and Figure \ref{fig:freq_sigma} (b)). These results prove the effectiveness of the proposed method to measure both thickness and electrical conductivity with a single frequency measurement. This is confirmed by the minimum values obtained for the errors and standard deviations of both thickness and conductivity that can reach values lower than 0.1 \%.

To complete the analysis for this first experimental validation, Tables \ref{tab:results_thickness} and \ref{tab:results_sigma} show the figures of merit on the estimation of electrical conductivity and thickness obtained by averaging the individual estimates at each angular frequency. This is a simple scheme that can be replaced by more advanced schemes combining all the level curves from different frequencies in the final estimate.

\begin{table}[htb]
\centering
\captionsetup{justification=centering}
\caption{Summary of the obtained results for the thickness estimation considering the overall executed tests.}
\begin{tabular}{ccccccc}
\hline \hline
Name code & $\widetilde{\Delta h}$ [mm] & $\overline{\Delta h}$ [mm] & $std_{\Delta h}$ [mm] & $\epsilon_{\Delta h}$ [\%] & $std_{\epsilon_{\Delta h}}$ [\%] \\ \hline
\#a       & 2.03                       & 2.10                      & 0.18              & 3.78     & 0.97                                  \\
\#b       & 0.984                       & 0.981                      & 0.091              & 0.40     & 1.34                                        \\
\#c       & 1.97                       & 1.98                      & 0.17              & 0.51        & 0.71                                        \\
\#d       & 1.03                       & 1.02                      & 0.14             & 0.91       & 1.45                                    \\
\#e       & 2.93                       & 2.90                      & 0.20              & 0.96      & 1.55                                         \\
\#f       & 3.98                       & 3.87                     & 0.19              & 2.77     & 0.97                                        \\
\hline \hline
\label{tab:results_thickness}
\end{tabular}
\end{table}

\begin{table}[htb]
\centering
\captionsetup{justification=centering}
\caption{Summary of the obtained results for the electrical conductivity estimation considering the overall executed tests.}
\begin{tabular}{ccccccc}
\hline \hline
Name code & $\widetilde{\sigma}$ [MS/m] & $\overline{\sigma}$ [MS/m] & $ std_{\sigma}$ [MS/m] & $\epsilon_{\sigma}$ [\%] & $ std_{\epsilon_{\sigma}}$ [\%]  \\ \hline
\#a       & 17.66                       & 17.27                      & 2.3              & 2.18  & 0.86                                        \\
\#b       & 58.50                       & 58.44                      & 6.8             & 0.10  & 2.2                                      \\
\#c       & 28.23                      & 28.26                      & 4.8             & 0.11  & 1.1                                          \\
\#d       & 35.27                      & 35.77                      & 4.2              & 1.44    & 2.1                                        \\
\#e       & 35.44                      & 35.33                      & 2.1             & 0.31    & 1.1                                          \\
\#f       & 35.91                       & 36.76                   & 2.3              & 2.38   & 0.86                                           \\
\hline \hline
\label{tab:results_sigma}
\end{tabular}
\end{table}

Finally, Table \ref{tab:comparison} shows a comparison between the performance levels achieved in different methods. Some methods provide a single estimate of either thickness or electrical conductivity, while other methods provide the estimate of both quantities. The table shows the range of thicknesses and/or electrical conductivities investigated in the related paper, together with the metrological performance levels expressed in terms of average relative error ($\overline{\epsilon}_{\Delta h}$, $\overline{\epsilon}_{\sigma}$) and maximum relative error ($\epsilon_{\Delta h}^{max}$, $\epsilon_{\sigma}^{max}$). From Table \ref{tab:comparison}, it follows that the proposed method outperforms all the other methods estimating both electrical conductivity and thickness \cite{MFD_double_est2,Comparison,Phase_double_est,MFD_double_est1}, while being in line with the performance levels of those methods retrieving one single parameter only \cite{Peyton1,Sardellitti,Sardellitti2,Sardellitti3,9815314,8466798}.

\begin{table}[htb]
\captionsetup{justification=centering}
\caption{Estimation performance comparison between the proposed method and some methods proposed in the scientific literature.}
\begin{adjustbox}{width=\columnwidth,center}
\begin{tabular}{ccccccc}
\hline \hline
                                                     & \multicolumn{2}{c}{Estimated parameters and analyzed ranges}                & \multicolumn{4}{c}{Estimation performance}                                                                          \\ \hline
Method  & Thickness ($\Delta h$) [mm]     & Electrical conductivity ($\sigma$) [MS/m] & $\overline{\epsilon}_{\Delta h}${[}\%{]} & $\epsilon_{\Delta h}^{max}${[}\%{]} & $\overline{\epsilon}_{\sigma}${[}\%{]} & $\epsilon_{\sigma}^{max}$ {[}\%{]}\\ \hline
Yin \textit{et al.} \cite{Peyton1}         & 1 ÷ 5                    & N/A                   & 0.88                          & 1.48                               & \multicolumn{2}{c}{N/A}           \\
    Sardellitti \textit{et al.} \cite{Sardellitti} & \multirow{2}{*}{0.5 ÷ 4} & N/A                   & 1.59                          & 3.19                               & \multicolumn{2}{c}{N/A}           \\
Sardellitti \textit{et al.} \cite{Sardellitti1} &                                        & N/A                   & 1.10                          & 2.43                               & \multicolumn{2}{c}{N/A}           \\
Sardellitti \textit{et al.} \cite{Sardellitti3} &       1 ÷ 4                                   & N/A                   & 1.26                          & 2.28                               & \multicolumn{2}{c}{N/A}           \\
Xie \textit{et al.} \cite{9815314} &       N/A                                   & 0.58 ÷ 58.2                    & \multicolumn{2}{c}{N/A}                             & 1.05 & 1.8           \\
Wang \textit{et al.} \cite{8466798} &       N/A                                   & 14.5 ÷ 51.5                   & \multicolumn{2}{c}{N/A}                             & 1.00 & 2.79           \\
Lee \textit{et al.} \cite{MFD_double_est2} &             1 ÷ 3.5                           & 0.56                 & 4.69                          & 8.72                               &  1.87     &   {N/A}  \\
Huang \textit{et al.} \cite{Comparison}       & 0.2 ÷ 1.5                & 0.6 ÷ 58           & 3.16                          & 6.70                               & 3.81                & 4.40                     \\ 
Huang \textit{et al.} \cite{Phase_double_est}       & 1 ÷ 4                & 0.6 ÷ 1.3           & 3.70                          & 5.40                              & 3.68               & 5.20                    \\ 
Lin \textit{et al.} \cite{MFD_double_est1}       & 0.4 ÷ 2.6                & 18.1 ÷ 58.9           & 3.89                          & 17.25                              & 5.74               & 21.14                    \\ \hline
Proposed method                                      & 1 ÷ 4                    & 17.7 ÷ 58.5         & 1.55                          & 3.78                               & 1.09                & 2.38                     \\ \hline \hline
\label{tab:comparison}
\end{tabular}
\end{adjustbox}
\end{table}

\section{Conclusions} \label{section:Conclusion}
This paper introduces dimensional analysis in Non--Destructive Testing \& Evaluation problems for the first time. 

The use of dimensional analysis makes it possible to represent a physical system via a minimal set of variables, the so-called \textpi \ groups. The \textpi \ groups are dimensionless quantities. Moreover, they can be found from the knowledge of only the physical dimensions of the variables describing the original problem. This makes it possible \textit{(i)} to reduce the computational cost when numerically simulating the physical system of interest, as required for quantitative inversion of the data, training of Artificial Intelligence algorithms, repeated simulations to design a new probe, etc., and \textit{(ii)} to represent an inverse problem in a reduced dimensional space, as in our application where a simple inverse problem was represented in a plane. 

In order to present the approach in as clear a manner as possible, an ECT method was proposed for the simultaneous estimation of the thickness and the electrical conductivity of conductive plates. The method has been presented from the underlying concept to a successful experimental validation carried out on metallic plates with different thicknesses (from 1 to 4 mm) and electrical conductivities (from 17 to 58 MS/m). The method can be applied to either single or multi-frequency data and its negligible computational cost makes it suitable for industrial in-line inspections.

From a general perspective, the proposed approach is highly effective for treating \emph{all} NDT problems where the number of unknowns is limited, as in many practical situations. Future work includes a complete metrological characterization of the proposed method, its extension to magnetic materials, where the key parameters are $\sigma$, $\Delta h$ and $\mu$, the optimization of the frequency range for the inspection and introduction of multi-frequency excitation signals.

\appendix
\section{Construction of the dimensionless groups}
\label{app:construction}
Here the method to compute the dimensionless group is outlined. We refer to \cite{Ciulla} and \cite{Tang} for details.

From a general perspective, each \textpi \ group can be expressed as
\begin{equation}
    \pi_i = q_1^{\alpha_{i1}} \cdots q_j^{\alpha_{ij}} \cdots q_n^{\alpha_{in}} 
    \label{eq:buck3}
\end{equation}
where the exponents $\alpha_{ij}$, $j = 1,2,\ldots,n$, are rational numbers such that $\pi_i$ is a dimensionless quantity.

To find the $\alpha_{ij}$s, a set of $k$ so called \lq\lq repeating variables\rq\rq \ (see \cite{Tang}), chosen among the $n$ dimensional quantities $q_1,\ldots,q_n$, has to be defined. The repeating variables must satisfy the following constraints: \textit{(i)} their products, with proper exponents, provide all the physical dimensions of the underlying problem; \textit{(ii)} they are independent; \textit{(iii)} their arbitrary nontrivial products do not generate a dimensionless quantity; \textit{(iv)} they should not be dependent variables of the problem, if any. Assuming the $k$ repeating variables are $q_1,\ldots,q_k$, each \textpi \ group is expressed as
\begin{equation}
    \pi_i=q_1^{\alpha_{i1}} \times \ldots \times q_1^{\alpha_{ik}} \, q_{k+i}, \ i=1,\ldots,n-k,
\end{equation}
and coefficients $\alpha_{ij}$s are found by imposing each \textpi \ group to be dimensionless.

As an example of application, the procedure for the construction of the dimensionless groups for the simple RLC series circuit in Figure \ref{fig:RLC_Circuit}, carrid out in the frequency domain, is proposed. The physical quantities are $n=6$ and listed in Table \ref{tab:RLC circuit}.

\begin{table}[htb]
\centering
\captionsetup{justification=centering}
\caption{Dimensional variables in the RLC circuit, expressed in terms of fundamental dimensions.}
\begin{tabular}{cccc}
\hline \hline
Number & Parameter            & Symbol  & Fundamental dimensions   \\ \hline
1      & angular frequency     & $\omega$ & {[}$A^0$ $V^0$ $T^{-1}${]} \\
2      & voltage               & $\Bar{E}$  & {[}$A^0$ $V^1$ $T^0${]}  \\
3      & electrical resistance & R        & {[}$A^{-1}$ $V^1$ $T^0${]} \\
4      & current               & $\Bar{I}$  & {[}$A^1$ $V^0$ $T^0${]}  \\
5      & inductance            & L        & {[}$A^{-1}$ $V^1$ $T^1${]} \\
6      & capacitance           & C        & {[}$A^1$ $V^{-1}$ $T^1${]} \\
\hline \hline
\end{tabular}
\label{tab:RLC circuit}
\end{table}

The electrical current $\Bar{I}$ is assumed to depend on the other quantities, i.e.
\begin{equation}
    \Bar{I} = f \left( \Bar{E},R,L,C,\omega \right).
    \label{eq:RLC1}
\end{equation}

A set of possible repeating variables is ${R,\Bar{E},\omega}$. Indeed, \textit{(i)} the fundamental dimensions can be expressed in term of monomial products of the repeating variables (see Table \ref{tab:Products}), \textit{(ii)} the repeating variables are independent and \textit{(iii)} arbitrary nontrivial monomial products do not generate dimensionless quantities and \textit{(iv)} they are not dependent variables. Condition \textit{(iii)} can be checked algebraically. Indeed, this condition is satisfied if and only if the kernel of the matrix made by the coefficients of the fundamental dimensions for the repeating variables is $\{ \mathbf{0} \}$. In this specific case, the exponents of the fundamental dimensions for $\omega$, $\Bar{E}$ and $R$ give the matrix
\[\begin{bmatrix}
0 & 0 & -1\\
0 & 1 & 0\\
-1 & 1 & 0
\end{bmatrix},\]
that which is invertible and, therefore, its kernel is $\{ \mathbf{0} \}$.

\begin{table}[htb]
\centering
\captionsetup{justification=centering}
\caption{Fundamental dimensions from monomial products of the repeating variables.}
\begin{tabular}{cc}
\hline \hline
Product & Dimension   \\ \hline
$E/R$  & $\left[ A \right]$  \\
$E$  & $\left[ V \right]$  \\
$\omega$  & $\left[ T \right]$  \\
\hline \hline
\end{tabular}
\label{tab:Products}
\end{table}

The \textpi \ groups arising from these choices, are in the form
\begin{align}
   \pi_1 & = R^{\alpha_{11}} E^{\alpha_{12}} \omega^{\alpha_{13}} I\\
   \pi_2 & = R^{\alpha_{21}} E^{\alpha_{22}} \omega^{\alpha_{23}} L\\
   \pi_3 & = R^{\alpha_{31}} E^{\alpha_{32}} \omega^{\alpha_{33}} C.
\end{align}

To compute coefficients $\alpha_{ij}$, it suffices to write a system of equations imposing that each \textpi \ group is dimensionless. For instance, with reference to $\pi_1$, this condition is
\begin{equation}
    A^0V^0T^0=\left( A^{-1}V^1T^0 \right) ^{\alpha_{11}} \left( A^0V^1T^0 \right) ^{\alpha_{12}} \left( A^0V^0T^{-1} \right) ^{\alpha_{13}} \left( A^1V^0T^0 \right),
\label{eq:RLC3}
\end{equation}
which gives $\alpha_{11}=1$, $\alpha_{12}=-1$ and $\alpha_{13}=0$ as solution of \eqref{eq:System}.

\begin{equation}
    \begin{cases}
    0 & = -\alpha_{11} + 1 \\
    0 & = +\alpha_{11} + \alpha_{12} \\
    0 & = -\alpha_{13}
    \end{cases}
    \label{eq:System}
\end{equation}

The complete list of \textpi \ groups is given in \eqref{eq:pi_groups_RLC}.

\printbibliography

\end{document}